\documentclass[aip,jcp,reprint]{revtex4-1}
\usepackage{graphicx,amsmath,amsfonts,amssymb,amsthm}
\begin{document}

\title[Adaptive two-regime method: application to front propagation]{Adaptive
two-regime method: application to front propagation}

\author{Martin Robinson}
\email{martin.robinson@maths.ox.ac.uk.}

\author{Mark Flegg}%
\email{mark.flegg@maths.ox.ac.uk.}
 
\author{Radek Erban}%
\email{erban@maths.ox.ac.uk.}

\affiliation{ 
Mathematical Institute, University of Oxford,
Andrew Wiles Building, Radcliffe Observatory Quarter,
Woodstock Road, Oxford OX2 6GG, United Kingdom
}%

\date{\today}

\begin{abstract}
The Adaptive Two-Regime Method (ATRM) is developed for hybrid (multiscale)
stochastic simulation of reaction-diffusion problems. It efficiently couples detailed 
Brownian dynamics simulations with coarser lattice-based models. The ATRM is 
a generalization of the previously developed Two-Regime Method [Flegg et al, 
Journal of the Royal Society Interface, 2012] to multiscale problems 
which require a dynamic selection of regions where detailed Brownian 
dynamics simulation is used. Typical applications include a front
propagation or spatio-temporal oscillations. In this paper, the ATRM is
used for an in-depth study of front propagation in a stochastic reaction-diffusion
system which has its mean-field model given in terms of the Fisher equation
[Fisher, Annals of Eugenics, 1937]. It exhibits a travelling reaction front 
which is sensitive to stochastic fluctuations at the leading edge 
of the wavefront. Previous studies into stochastic effects on the 
Fisher wave propagation speed have focused on lattice-based models, 
but there has been limited progress using off-lattice (Brownian dynamics) 
models, which suffer due to their high computational cost, particularly 
at the high molecular numbers that are necessary to approach the 
Fisher mean-field model. By modelling only the wavefront itself 
with the off-lattice model, it is shown that the ATRM leads to 
the same Fisher wave results as purely off-lattice 
models, but at a fraction of the computational cost. The error analysis
of the ATRM is also presented for a morphogen gradient model.
\end{abstract}


\maketitle

\section{Introduction}

\noindent
Deterministic mean-field models of reaction-diffusion systems describe the state 
of each chemical species using continuous variables (concentrations) and simulate
its variation with time and space using reaction-diffusion partial differential
equations (PDEs). Due to the wealth of analytical techniques available, such
mean-field models have enjoyed considerable success \cite{Murray:2002:MB}. 
However, there has been increasing interest in modelling the stochastic effects 
that arise from either a finite population size or the discrete nature of its 
individuals \cite{Black:2012:SFE,Gillespie:2013:PSA}.
Stochastic models are often derived using a bottom-up approach, where the model
is formulated using the individuals of the population as the basic entities, and
the model parameters are chosen so that in the limit of large population size
the model approaches known mean-field diffusion and reaction rates
\cite{Andrews:2004:SSC,Lipkova:2011:ABD,Engblom:2009:SSR}. This type
of model is sometimes termed an individual-based model (IBM) 
\cite{Black:2012:SFE}.

Generally, different IBMs can be divided into one of two separate categories:
off-lattice or lattice-based models. Off-lattice models treat each individual as
a point in a continuous spatial domain. Different individuals are more likely to interact if they are located in
a similar spatial location, i.e. the likelihood of interaction
often depends on the distance between each pair \cite{Erban:2009:SMR}. 
Here we restrict our consideration to the diffusion and reaction of molecular 
species and use the term molecular-based instead of off-lattice model. 
Molecular-based simulations in this paper are formulated in the form of 
Brownian dynamics \cite{vanGunsteren:1982:ABD,Lipkova:2011:ABD}.
Each molecule of each species is given a position in the spatial domain, and 
bimolecular reactions can occur whenever two molecules are separated by 
a given binding radius \cite{Smoluchowski:1917:VMT}. 

Lattice-based models involve the discretization of the computational domain into
a set of compartments (i.e. a lattice), upon which individuals can move by ``jumping'' 
between neighbouring compartments (i.e. connected lattice sites) \cite{Erban:2007:PGS}.
In the applications which we shall consider, the
individuals (molecules) do not have memory, that is, they do not remember which
lattice site they came from. Therefore, compartment-based models are
particularly suitable for efficient simulations, as only the number of molecules
in each compartment is recorded. For this type of model the concept of a lattice
has been replaced with a set of connected compartments, each with a specific
volume. This volume, and the molecules within it, are assumed to be well-mixed.
Reactions can only occur between molecules in the same compartment, and
diffusion occurs by random jumps between neighbouring compartments
\cite{Erban:2007:PGS,Erban:2009:SMR}.

\begin{figure}[htbp]
\begin{center}
  \includegraphics[width=0.44\textwidth]{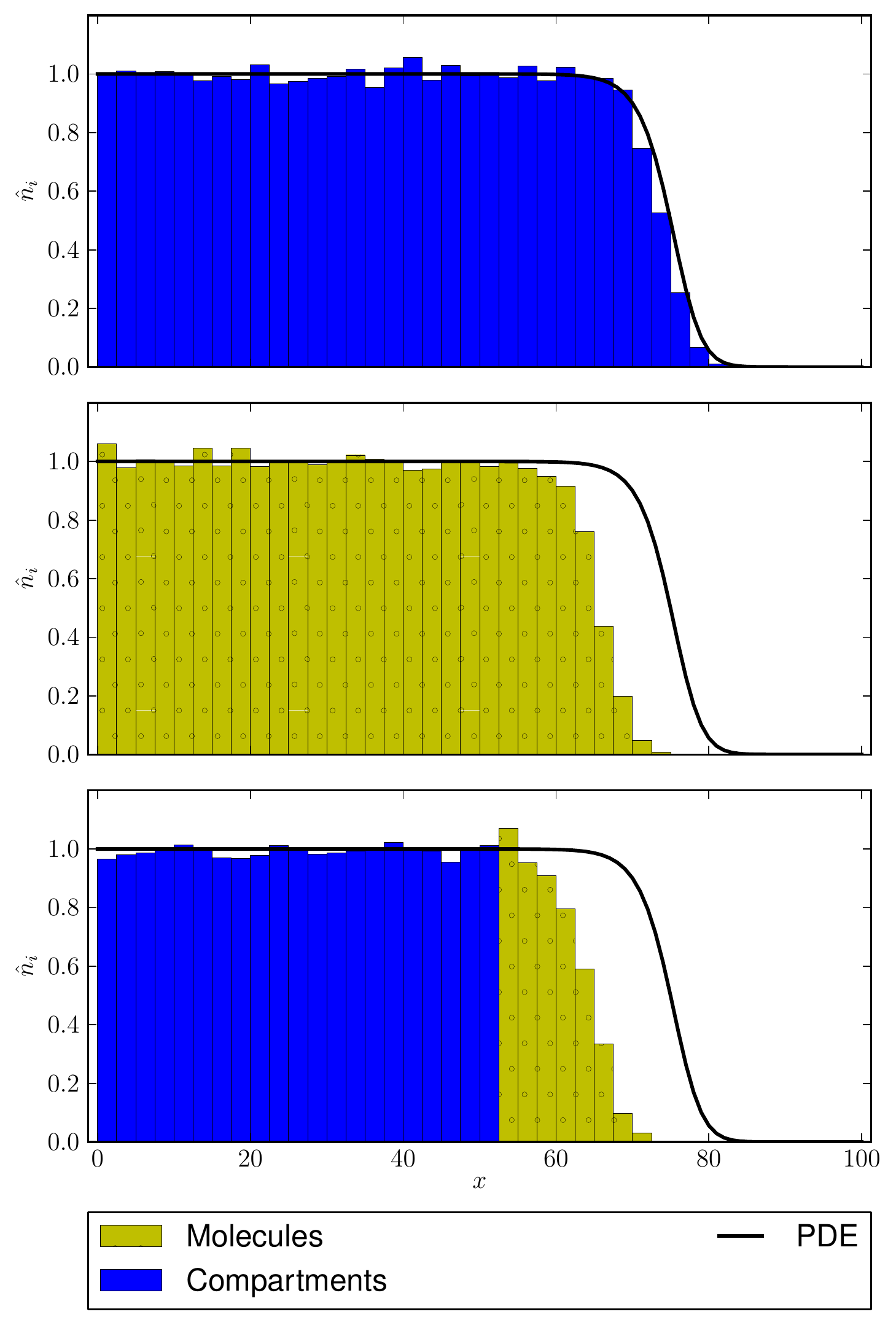}
  \caption{{\it Time snapshots of the profile of the stochastic Fisher wave
  (see Section {\rm \ref{sec:fisher_wave}} for details) using different IBMs,
  demonstrating the differences in wave speed between the models. 
  Top: Compartment-based simulation (Section {\rm\ref{sec:compartment_method}}); 
  Middle: Molecular-based simulation
  (Section {\rm\ref{sec:molecular_method}}); Bottom: Coupled molecular-based and
  compartment-based simulation (Section {\rm\ref{sec:two_regime_method}}). Plots
  show a histogram of scaled molecular concentration versus the $x$-coordinate.
  Boxes colored blue are compartment-based regions while yellow (with little
  circles) denotes a molecular-based region.
  The blue line shows a (deterministic) finite difference solution of
  the mean-field PDE $(\ref{eq:fisher_eq})$.}}
  \label{fig:default_fisher_sim}
\end{center}
\end{figure}

Different IBMs can exhibit different stochastic effects, even for large
molecule numbers\cite{Black:2012:SFE}. In particular, a well known
reaction-diffusion system that exhibits a slow convergence to the mean field
description is the Fisher travelling wave, the prototype model for the spread of
a biological species\cite{Fisher:1937:WAA}. For finite molecule numbers,
stochastic effects can play a significant role in reducing the speed of the
wave\cite{Panja:2004:EFP}, but the degree to which the wave speed is reduced
depends on the details of the particular IBM used.
Figure \ref{fig:default_fisher_sim} shows snapshots from Fisher wave simulations
using both compartment-based (top plot) and a molecular-based (middle plot)
IBMs. The parameters and initial conditions are the same for both simulations
and are discussed in Section
\ref{sec:fisher_wave}. It is evident that one obtains substantially different
wave speeds depending on the IBM used, even though they would be both described by the same mean-field model.
The solution of mean-field PDE (\ref{eq:fisher_eq}) is shown as a solid
blue line. Note that the mean-field model does not include any
stochastic effects and thus should not be compared with the stochastic models in
terms of solution accuracy. While the compartment-based IBM shown here is much closer to the mean-field model than
the molecular-based IBM, this is merely an effect of the parameters used, and the
results in Section \ref{sec:fisher_wave} show that the compartment-based wave
speed is strongly dependent on the compartment size.

While there have been numerous investigations into the speed of Fisher
travelling waves for lattice or compartment-based IBMs
\cite{Panja:2004:EFP,vanSaarloos:2003:FPU,Brunet:2001:EMN,Breuer:1994:FEW},
molecular-based models have received little attention due to their heavy
computational requirements, which scale up quickly as the total number of
molecules increases.
An alternative and more efficient approach is to simulate only the wavefront
itself with a molecular-based IBM, while a less computationally intensive IBM is
used in the remainder of the computational domain, particularly in the high 
concentration region behind the wavefront. This ensures that the dynamics of 
the wave are determined
solely by the molecular-based IBM while keeping the total number of discrete
molecules (and the computational requirements) to a minimum. An example of this
(taken from the results shown in Section \ref{sec:fisher_wave}) is shown in
Figure \ref{fig:default_fisher_sim} (bottom), which matches the wave speed of
the purely molecular-based simulation.

A related hybrid approach to stochastic Fisher wave simulation was taken by Moro
\cite{Moro:2004:HMS}, who used a lattice-based model for the leading edge of
the wavefront and a numerical approximation of the mean-field equation for the
remainder of the domain. Moro's aim was also to use a hybrid approach for
computational efficiency, but was primarily interested in the lattice-based
model for the Fisher wave. In contrast, our goal is to study the effects of
using a molecular-based model of the Fisher wave, and we therefore use this
method to simulate the wavefront.

In this paper, we develop the Adaptive Two-regime Method (ATRM) which is suitable
for the efficient modelling of reaction-diffusion systems using both
molecular-based and compartment-based IBMs. This is achieved by the coupling 
of the two different stochastic models across an interface separating two 
respective non-overlapping spatial regions.
The ATRM is the generalization of the previously developed Two-Regime Method
(TRM) \cite{Flegg:2012:TRM,Flegg:2013:ATM} which allows a model to use 
different IBMs in regions in which they are required. One of the limitations 
of the TRM is that it has used a fixed domain decomposition. 
The ATRM generalizes TRM to any problem where the interface between the 
two regions can move over time, and therefore adapt to a dynamical chemical 
system (for example, travelling waves, regional population growth) or changes 
in the problem geometry (for example, cellular morphology). A simplified
version of ATRM was used previously to simulate the growth of filopodia 
\cite{Erban:2013:MSR}, but the focus was on the application itself rather 
than the error introduced by the moving interface. The goal of this paper 
is to both fully characterise the moving interface error for three-dimensional 
reaction-diffusion simulations, and to demonstrate that it can reproduce 
the results of a much more computationally demanding molecular-based method 
when applied to a Fisher wave simulation.

The paper is divided into three main sections. In Section \ref{sec:methods}, the
different methods used in this paper are described. In Section
\ref{sec:morphological_gradient}, we investigate the error introduced by the
moving interface and how it varies with the model parameters by applying the
method to a steady-state morphological gradient problem. These results are
compared to both a static interface and purely molecular-based and purely
compartment-based IBMs. Section \ref{sec:fisher_wave} describes the application
of the ATRM to a Fisher wave.
The speed of the modelled wave (corresponding to the interface speed) and the
effect of the total number of molecules and the compartment size are
investigated and compared with the purely molecular and compartment-based IBMs.
We conclude our paper with discussion of our results and other hybrid
(multiscale) methods for reaction-diffusion processes in Section \ref{secdiscussion}.

\section{Methods} \label{sec:methods}

\noindent
The most detailed modelling approach considered in this paper will be
given in terms of Brownian dynamics and introduced in Section 
\ref{sec:molecular_method}. We will use this model to study a Fisher wave
in Section \ref{sec:fisher_wave}. The rest of the models introduced below
will be used to decrease the computational intensity of the molecular-based
model, while keeping the same level of accuracy. 

\subsection{Molecular-based Modelling}
\label{sec:molecular_method}

\noindent
We will study a time-driven molecular-based algorithm in this paper 
\cite{Andrews:2004:SSC}.
Time-driven molecular-based algorithms consider each molecule as a single point
particle with position $\mathbf{x}(t)$. The molecular-based method
proceeds with discrete timesteps $\Delta t$, and the diffusion of the
molecules/particles is modelled as a discretized Brownian motion
\begin{equation}
\mathbf{x}(t + \Delta t) = \mathbf{x}(t) + \sqrt{2 D \Delta t}
\, \mathbf{\zeta},
\end{equation}
where $D$ is the diffusion constant and
$\mathbf{\zeta}=[\zeta_x,\zeta_y,\zeta_z]$ is a vector of random numbers sampled
from a normal distribution with zero mean and unit variance.
Examples of software packages implementing a time-driven molecular-based method
include Smoldyn \cite{Andrews:2004:SSC,Andrews:2012:SSC} and MCell
\cite{Stiles:2001:MCM,Kerr:2008:FMC}.

The simulation of zeroth-order reactions (production from a source with a fixed rate)
and first-order (unimolecular) reactions is relatively straightforward
and makes use of a generator of Poisson and exponentially distributed
random numbers \cite{Andrews:2004:SSC,Erban:2007:PGS}.
Bimolecular reactions can occur whenever two reactant molecules 
come within a specified distance of each other. In the Fisher wave 
simulation in Section \ref{sec:fisher_wave}, we have the following 
reversible bimolecular reaction
\begin{equation}
A + A
\underset{k_1}{\overset{k_2}{\rightleftarrows}} 
A
\label{reversiblereaction}
\end{equation}
with forward and backwards reaction rate constants $k_2$ and $k_1$,
respectively.
To model it, we follow a generalization to the classical Smoluchowski model, 
where the forward reaction occurs within the binding radius $\rho$ with
probability $P_{\Delta t}$ per timestep \cite{Erban:2009:SMR}. We also introduce an
unbinding radius $\alpha \rho$, which is the distance that the two molecules of $A$ 
are placed apart, whenever the backward reaction in
(\ref{reversiblereaction}) occurs \cite{Andrews:2004:SSC}.
To calculate $P_{\Delta t}$ and $\alpha$, we introduce the following dimensionless 
parameters 
$$
\gamma = \frac{\sqrt{4 \, D \,\Delta t}}{\rho}, 
\quad 
\kappa = \frac{k_{2} \Delta t}{\rho^3},
$$
where $D$ is the diffusion constant of $A$. The reaction probability per 
timestep $P_{\Delta t}$ can be found by solving (via a look-up table or root 
finding method)\cite{Lipkova:2011:ABD}
\begin{equation}
\kappa = 2 \pi P_{\Delta t} \int_0^1 \xi^2 g(\xi;P_{\Delta t},\gamma) \ \mathrm{d}\xi,
\label{calcro}
\end{equation}
where $g(\xi;P_{\Delta t},\gamma)$ is
found by discretizing and solving numerically
\begin{align*}
\begin{split}
g(r) = (1-P_{\Delta t}) \int_0^1 K(r,r';\gamma) g(r') 
\;\mbox{d}r' \\
+ \int_1^{\infty} K(r,r';\gamma) g(r') \;\mbox{d}r' \\ 
+ \frac{P_{\Delta t} K(r,\alpha;\gamma)}{\alpha^2} \int_0^1 g(r')r'^2
\;\mbox{d}r'.
\end{split}
\end{align*}
where 
\begin{equation*}
K(\xi,\xi';\gamma) = (4\pi \gamma P_{\Delta t})^{-1/2}
\exp\left(\frac{-(\xi-\xi')^2}{4\gamma P_{\Delta t}}\right)
\end{equation*} 
is Green's function for the diffusion PDE.

\subsection{Compartment-based Modelling}\label{sec:compartment_method}

\noindent
The domain is partitioned into $K$ compartments $j = {0,1,\ldots,K}$. Whilst
there has been significant progress in the field of irregular lattice
compartment-based reaction-diffusion 
simulation\cite{Engblom:2009:SSR,Drawert:2012:UMF}, here we restrict the
partitioning to a regular grid of cube compartments with side length $h$. The
molecules within each compartment are assumed to be well-mixed and are therefore
evenly distributed over its volume. Without the need for position information,
this method only stores the total number of each species contained
within each compartment. In this paper, all models will only include
one chemical species, $A$. We will denote the number of molecules of $A$
in the $j$-th compartment as $A_j$.

The compartment-based algorithm is event-based. In this paper, we use
a variant of the Next Subvolume Method \cite{Elf:2004:SSB} which is
itself an extension of the Gillespie algorithm \cite{Gillespie:1977:ESS} 
and the Gibson-Bruck algorithm \cite{Gibson:2000:EES}. At the beginning
of the simulation, the next event time $t_j$ is generated for each 
compartment by  
\begin{equation} \label{eq:tau}
t_j = \frac{1}{\alpha_j} \ln \left ( \frac{1}{u_0} \right ),
\end{equation}
where $u_0$ is a uniformly distributed random number in $(0,1)$ and
$\alpha_j$ is the sum of propensities of all events (reactions or
diffusion jumps) which can occur in the $j$-th compartment. In the Fisher wave 
simulation in Section \ref{sec:fisher_wave}, we have two reactions in each 
compartment, given as the forward and backward reactions in (\ref{reversiblereaction}).
Diffusion events (instantaneous jumps from a compartment to an adjacent 
compartment) are considered as reaction events \cite{Erban:2007:PGS}, 
with a propensity $D/h^2 A_{j}$.
Thus $\alpha_j = k_1 A_j + k_2 A_j (A_j-1) + 6 D/h^2 A_j$ in (\ref{eq:tau})
for the internal compartments, i.e. the compartments which have six neighbouring
compartments. Boundary compartments have appropriately modified propensity 
functions because they have less neighbouring compartments \cite{Erban:2007:RBC}.
Then the compartments are sorted by $t_j$ using an index priority queue. 
At each step of the algorithm the compartment $j_1$ with the smallest 
next reaction time is taken from from the queue and an event is chosen 
using another uniformly distributed random number \cite{Gillespie:1977:ESS}.
This event is processed and a new $t_{j_1}$ is
sampled for that compartment using
$$
t_{j_1} = t + \frac{1}{\alpha_j} \ln \left ( \frac{1}{u_0} \right ),
$$
where $t$ is the current time. If the processed event is a diffusion jump to
compartment $j_2$, then $A_{j_2}$ also changes and the corresponding $\alpha_{j_2}$ needs
to be recalculated. Denoting its old value as $\alpha^{old}_{j_2}$,
the old next event time $t_{j_2}^{old}$ is updated using \cite{Gibson:2000:EES}
\begin{equation}
t_{j_2} = t + \frac{\alpha^{old}_{j_2}}{\alpha_{j_2}} 
(t_{j_2}^{old} - t).
\end{equation}
Examples of software packages that implement the compartment-based model are
MesoRD \cite{Hattne:2005:SRD} and URDME \cite{Drawert:2012:UMF}.

\subsection{Two-Regime Method}\label{sec:two_regime_method}
 
\noindent  
The Two-Regime Method (TRM) was originally presented in one
spatial dimension \cite{Flegg:2012:TRM} and later extended to higher-dimensional 
domains \cite{Flegg:2013:ATM}. It considers the diffusion of molecules across 
the interface $I$ between non-overlapping domains $\Omega_C$ and
$\Omega_M$ modelled using compartment-based ($\Omega_C$) and molecular-based
methods ($\Omega_M$), respectively.

The TRM optimally preserves the correct diffusion flux across the interface $I$ 
between the regimes. To achieve this, a number of different factors must be
taken into account. When particles cross the interface into the compartment
domain $\Omega_C$ they are placed in a compartment. Whilst these molecules would 
ordinarily
be close to the interfacial side of the compartments in which they are placed, by
virtue of being described using a compartment-based approach, they must be
considered indistinguishable from other molecules ``spread out'' over the
compartment volume. In order to counterbalance the generated net flux from the
molecular-based domain $\Omega_M$ into the compartment-based regime $\Omega_C$
as a result of this
paradigm-critical loss of information, the propensity
of a diffusion jump back across the interface $I$ is specified differently 
to the other diffusive jump propensities using \cite{Flegg:2012:TRM}
$$
\frac{2h}{\sqrt{\pi D \Delta t}} \frac{D}{h^2} A_{j},
$$
where $A_j$ is the number of molecules in the compartment next to the interface $I$.
When a diffusion jump from $\Omega_C$ across the interface (to the molecular-based 
side $\Omega_M$) occurs
the molecule is given a position in $\Omega_M$ with a normal distance from the
interface given by $x$, where $x$ is sampled from \cite{Flegg:2012:TRM}
\begin{equation}\label{perpend}
f(x) = \sqrt{\frac{\pi}{4 D \Delta t}} \mathrm{erfc} \left (  \frac{x}{\sqrt{4
D \Delta t}} \right ).
\end{equation}
The perpendicular distance $x$ given by the distribution (\ref{perpend}) is
taken from an initial position on the interface given by $\mathbf{r}_j$
$$
\mathbf{r}_j = \mathbf{m}_j + y \mathbf{p}_1 + z \mathbf{p}_1,
$$
where $\mathbf{m}_j$ is the mid-point of the compartment face from which the
diffusion jump occurred, $\mathbf{p}_{1}$ and $\mathbf{p}_{2}$ are perpendicular
unit vectors tangential to the interface and aligned with the lattice vectors of the
compartment-based domain. Random numbers $y$ and $z$ are sampled from the triangular distribution
with lower limit $-h/2$, upper limit $h/2$ and zero mean
\cite{Flegg:2013:ATM}.

The TRM has been used previously to study filopodia dynamics
\cite{Erban:2013:MSR} and intracellular calcium release from ion channels
\cite{Flegg:2013:DSN} whereby a small-scale biochemical system is coupled with a
coarser model in a much larger domain.

\subsection{Adaptive Two-Regime Method} \label{sec:moving_interface}

\begin{figure}[htbp]
\begin{center}
  \includegraphics[width=0.5\textwidth]{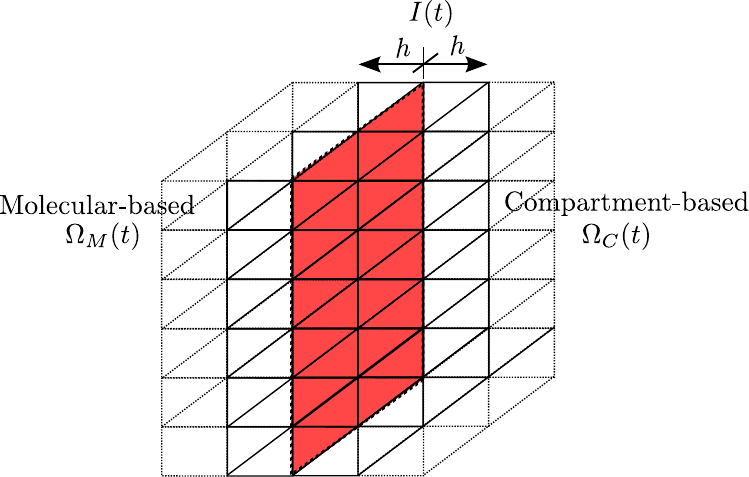}
  \caption{{\it The interface $I(t)$ between the molecular and compartment regions
  (red surface) is restricted to move by jumps between neighbouring compartment
  faces.}}
  \label{fig:moving_interface}
\end{center}
\end{figure}

\noindent In this paper we introduce the Adaptive Two-Regime Method (ATRM), a
method for changing the compartment and molecular-based subdomains $\Omega_C$
and $\Omega_M$ in response to the outcome and requirements of a dynamic
reaction-diffusion simulation. This is achieved by moving the interface $I
\equiv I(t)$ between simulation regimes.
Whilst the methodology introduced in this paper can be generalized for any
criteria defining the dynamic interface, we move the interface $I(t)$ in such a
way as to limit the computational requirements of the molecular-based subdomain
$\Omega_M$ (which can otherwise become too cumbersome). We will not be
considering time-adapting lattices in the compartment-based model and therefore
the moving interface $I(t)$ moves discretely such that the interface aligns with
the faces of the compartments.
The compartment geometry that we consider in this paper is a regular grid of
equal sized cubes with side length $h$ (see Figure \ref{fig:moving_interface}).
The interface $I(t)$ is constrained to move by step sizes equal to $h$ in a
direction normal to the interface surface, so that it is always flat and aligned
to the faces of those compartments on the boundary.

The conditions on the movement of the interface can be set according to the
specific problem. However, an obvious and useful goal is that the interface $I(t)$
moves to restrict the total number of particles in $\Omega_M$,
therefore placing a limit on the computational effort applied to the method that is
expected to be the most computationally demanding.

Following this goal, we move the interface towards the molecular-based region 
$\Omega_M$ if
the concentration of particles within a distance $h$ of the interface is above a
given threshold $c_{\mathrm{max}}$. Conversely, we move move the interface
towards the compartment region $\Omega_C$ if the concentration of molecules within the
compartments on the boundary is less than $c_{\mathrm{max}}-\delta c$. In this
way the maximum concentration expected in the molecular region is below
$c_{\mathrm{max}}$. The difference between the two thresholds $\delta c$ is
necessary to prevent the spurious movement of the interface due to stochastic
fluctuations in the molecule concentrations \cite{Ho:2012:MSR}. As we
shall see, the reduction of $\delta c$ to zero results in inaccuracies in the
simulation that are due to rapid redistribution of molecules that occurs every
time the interface $I(t)$ moves into $\Omega_M$. We check for the
upper and lower limit concentrations and move the interface every $n_c$
timesteps of the simulation. Checking this condition at every time step ($n_c =
1$) is unnecessary and can be computationally costly and therefore
counterproductive to the purpose of using this multiscale method \cite{Ho:2012:MSR}.

After each check of the movement criteria, the interface $I(t)$ can either move into
the molecular region, or into the compartment region. If the former, then each
molecule that is in the new compartment region (a
perpendicular distance $h$ from the old interface) is counted and placed inside
the corresponding newly created compartment $j_{new}$. For each particle that is
removed from the molecular-based simulation, the copy number in the new
compartment $A_{j_{new}}$ is incremented by one. If the interface moves into the
compartment region then each compartment $j$ which was previously adjacent to
the interface and now in the new molecular region is removed and $A_{j}$ new
molecules are created within the space occupied by the old $j$-th compartment with
randomly-generated, uniformly-distributed initial positions.

\section{Results}

\noindent
Two model problems will be considered in this paper. 
In Section \ref{sec:morphological_gradient} we demonstrate the impact 
of applying the ATRM to a simple morphogen gradient 
problem\cite{Tostevin:2007:FLP,Howard:2012:HBR} 
with a known solution. This allows for an easy comparison between 
simulations using both static and moving interfaces. In
this way, the error associated with the moving interface 
will be be studied. In Section \ref{sec:fisher_wave} we
use the ATRM to investigate Fisher waves in a molecular-based
model.

\subsection{Steady State Morphological
Gradient}
\label{sec:morphological_gradient}

\begin{figure}[htbp]
\begin{center}
  \includegraphics[width=0.5\textwidth]{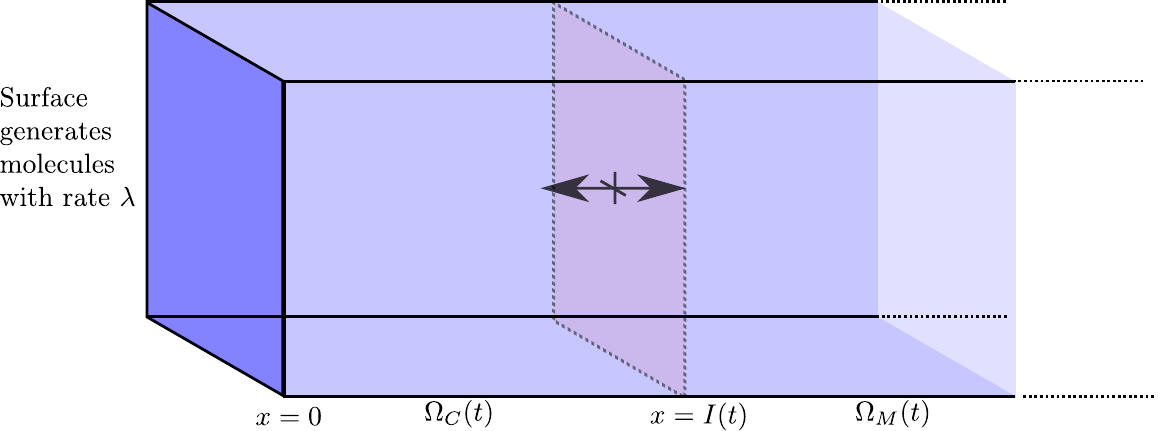}
  \caption{{\it The simulation domain $\Omega = (0,\infty)\times(0,1)\times(0,1)$.
  The moving interface is located at $x=I(t)$, and the compartment and molecular
  regions are $\Omega_C = (0,I(t))\times(0,1)\times(0,1)$ and $\Omega_M =
  (I(t),\infty)\times(0,1)\times(0,1)$ respectively.
  Molecules are generated at the $x=0$ boundary (coloured dark blue) with rate 
  $\lambda$. This
  boundary is also reflective. There is no lower boundary in the positive $x$
  direction. All other boundaries in the $y$ and $z$ directions are periodic.}}
  \label{fig:moving_interface_domain}
\end{center}
\end{figure}

\noindent
The simulation domain is a semi-infinite cuboid shown 
in Figure \ref{fig:moving_interface_domain}.
The boundary at $x=0$ (coloured dark blue) is reflective and generates molecules
with rate $\lambda$. There is no upper boundary in the $x$ direction and the
molecules are allowed to diffuse to $x \rightarrow \infty$. The
compartment-based and molecular-based subdomains are labelled $\Omega_C$ and
$\Omega_M$, respectively. The interface between the subdomains is a plane
perpendicular to the $x$-axis at $x=I(t)$ and moves parallel to the $x$-axis
with constant step size $h$. All boundaries in the $y$ and $z$ directions are
periodic. One species $A$ is simulated and moves with diffusion constant $D$.
In addition to the production of molecules at $x=0$, one unimolecular 
degradation reaction 
$$
A \overset{k}{\rightarrow} \emptyset
$$ 
is simulated. Thus, in the limit of high molecule copy numbers, the normal 
rate equations for this system give
$$
\frac{da(x,t)}{dt}=D\frac{\partial^2
a(x,t)}{\partial x^2} - ka(x,t) + \lambda \delta(x)
$$
where $a(x,t)$, $x \ge 0$, $t \ge 0$, denotes the concentration of $A$
at any point $(x,y,z) \in \Omega$.
This equation can be explicitly solved\cite{Bergmann:2007:PDB} to give 
\begin{align}\label{eq:morph_gradient_soln}
\begin{split}
a(x,t) = \frac{\lambda}{2 \beta D} \biggl [e^{-\beta x} - \frac{e^{-\beta
x}}{2} \mbox{erfc} \left (\frac{2 \beta D t-x}{\sqrt{4 D t}} \right)\\
 -
\frac{e^{\beta x}}{2} \mbox{erfc} \left (\frac{2 \beta D t+x}{\sqrt{4 D t}}
\right) \biggr ]
\end{split}
\end{align}
where $\beta = \sqrt{k/D}$.

\subsubsection{Transient regime}\label{sec:transient}

\noindent
The time-varying solution given in equation (\ref{eq:morph_gradient_soln}) has an
initial transient period leading to a steady state solution
\begin{equation}
a_s(x)=\frac{\lambda}{2 \beta D}  e^{-\beta x}
\label{steadystatea}
\end{equation} as $t \rightarrow \infty$.
This section examines the initial transient period, using the 
parameters given in Table \ref{tab:transient_sim_params}.

\begin{table}
\begin{center}
\begin{tabular}{|c | c|}
\hline
Parameter & Value \\
\hline
$D$ & 1 \\
$k$ & 10 \\
$\lambda$ & $10^6$ \\
$\beta = \sqrt{k/D}$ & $\sqrt{10}$ \\
\hline
$h$ & 0.05 \\
$\Delta t$ & $10^{-4}$ \\
\hline
$c_{max}$ & $a_s (1/2) = 0.206 \, a_s (0)$ \\
$\delta c$ & $0.02 \, a_s(0)$ \\
$n_c$ & 10 \\
\hline
\end{tabular}
\caption{{\it Table of simulation parameters for the morphological gradient simulation.
The first three parameters are the parameters of the biological model
($D$, $k$ and $\lambda$). Parameter $h$ is the compartment size in $\Omega_C$
and parameter $\Delta t$ is the time step in $\Omega_M$. The last three 
parameters $c_{max}$, $\delta c$ and $n_c$ are the parameters of the {\rm ATRM}.
The function $a_s$ is given by $(\ref{steadystatea})$.}}
\label{tab:transient_sim_params}
\end{center}
\end{table}

\begin{figure}[htbp]
\begin{center}
  \includegraphics[width=0.5\textwidth]{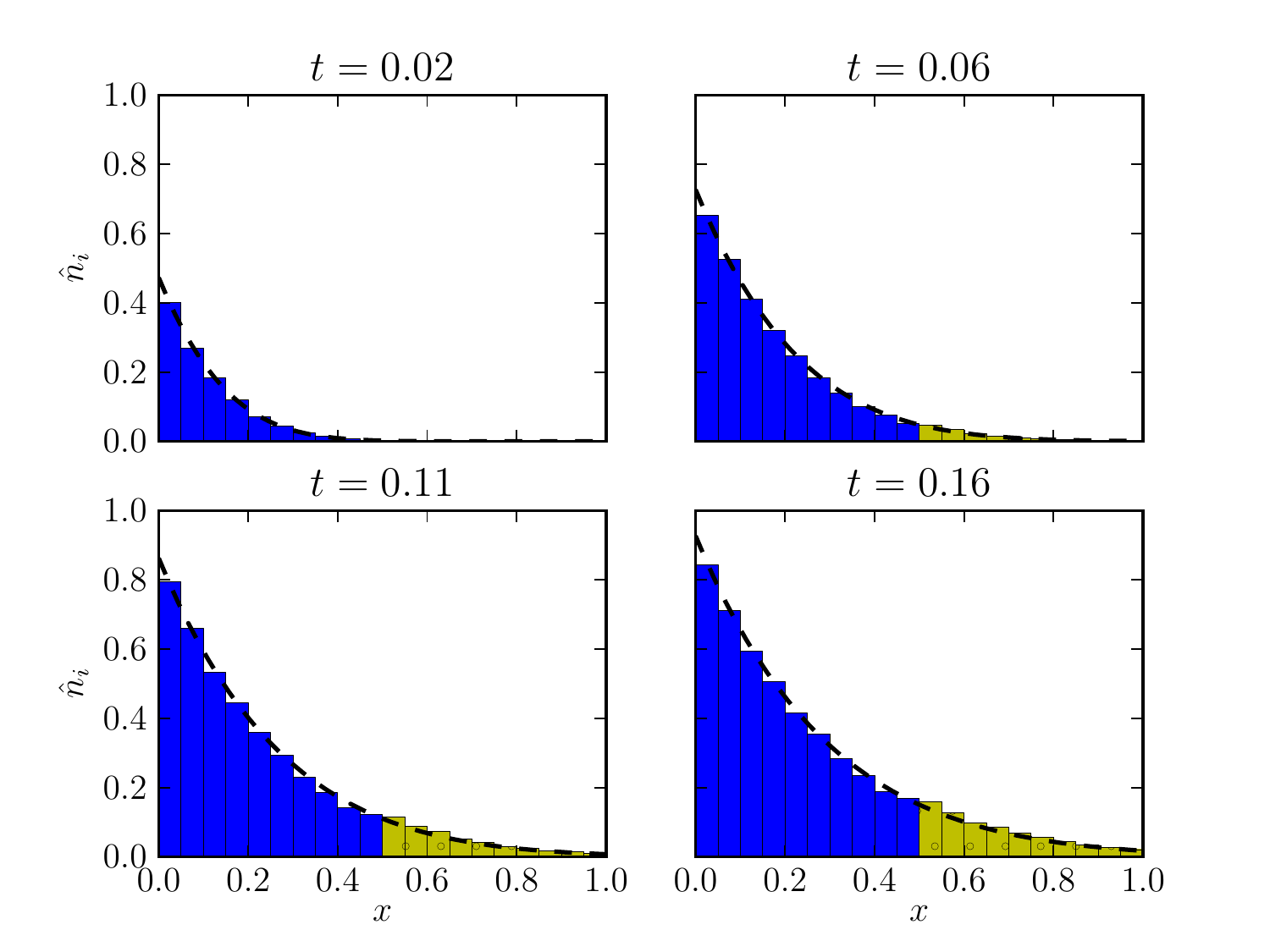}
  \caption{{\it Concentration histograms (along the $x$-axis) from the {\rm TRM}
  simulation of morphogen gradient with $I=0.5$. The concentration in each compartment
  $\hat{n}_i$ has been scaled by the maximum expected concentration $a_s(0)$.
  Blue denotes the compartment-based region while yellow (with little circles) is used
  for the molecular-based region.}}
  \label{fig:static_interface_concentration}
\end{center}
\end{figure}

\begin{figure}[htbp]
\begin{center}
  \includegraphics[width=0.5\textwidth]{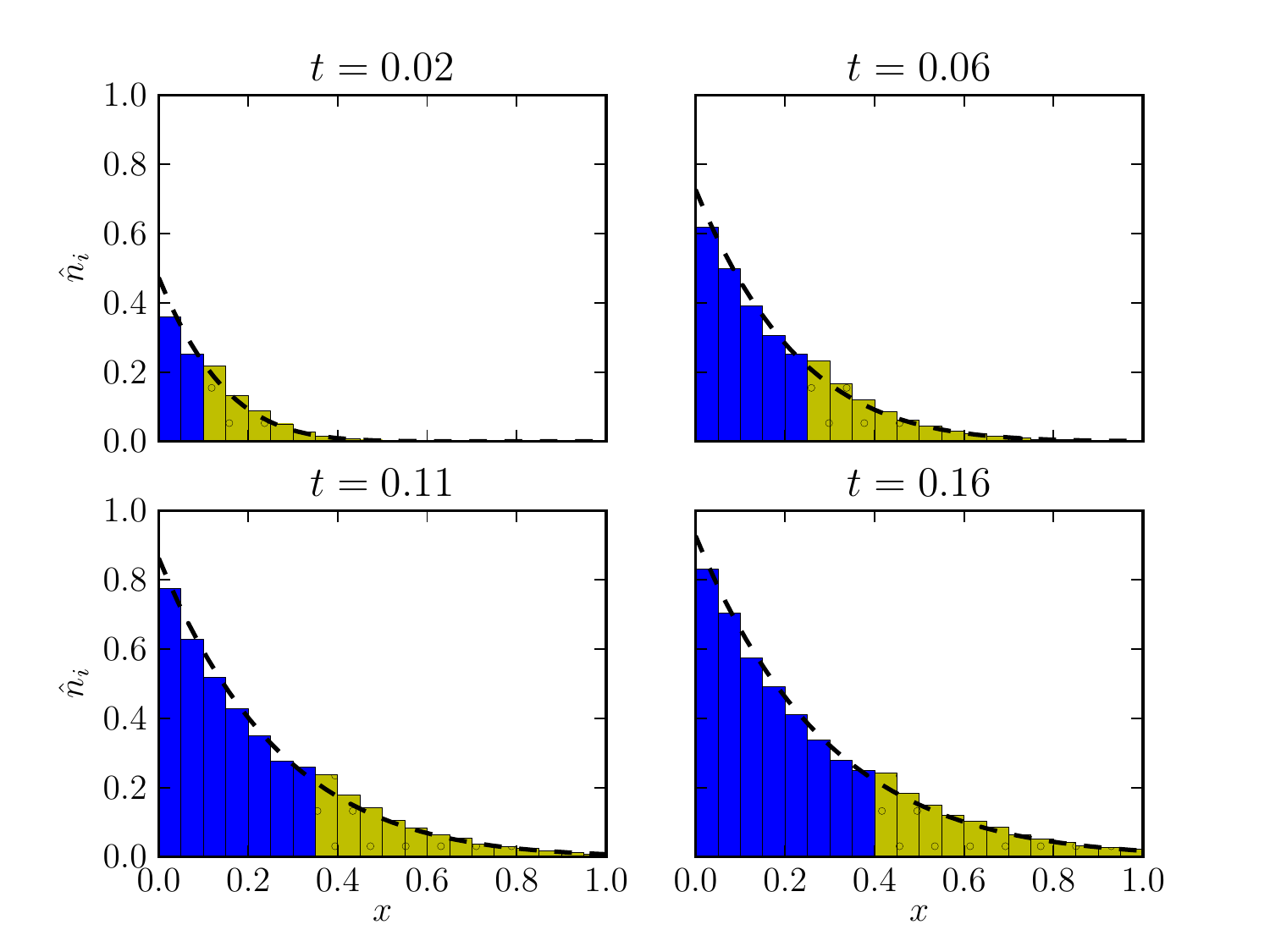}
  \caption{{\it Concentration histograms (along the $x$-axis) from the 
  {\rm ATRM} simulation of morphogen gradient with parameters given
  in Table {\rm\ref{tab:transient_sim_params}}. The concentration in each
  compartment $\hat{n}_i$ has been scaled by the maximum expected concentration $a_s(0)$.
  Blue denotes the compartment-based region while yellow (with little circles) is used
  for the molecular-based region.}}
  \label{fig:moving_interface_concentration}
\end{center}
\end{figure}

Figure \ref{fig:static_interface_concentration} shows the results from a
classical TRM simulation with a static interface $I(t) \equiv 0.5$, i.e. the bottom
three parameters in Table \ref{tab:transient_sim_params} are not used.
Four different timesteps (at $t = 0.02$, $0.06$, $0.11$ and
$0.16$) were chosen from the transient period and the data from the molecular
and compartment regions were plotted together in a one-dimensional concentration
histogram along the $x$-axis. The bin size of the histogram was chosen to match
the compartment size $h$. At all times shown, the concentration histogram data
shows a good agreement with the analytical solution $a(x,t)$ in equation
(\ref{eq:morph_gradient_soln}).
Figure \ref{fig:moving_interface_concentration} shows similar results but from
the ATRM simulation with a moving interface for 
parameters in Table \ref{tab:transient_sim_params}. The moving interface between the
molecular and compartment regions correctly follows the maximum threshold set at
$c_{max}=0.206 \, a_s(0)$, and no noticeable differences can be seen between the
static (TRM) and moving (ATRM) interface results.

To measure the error, we count the number of molecules $N(t)$ in the 
region $(0,0.5) \times (0,1)\times(0,1)$ and compare this to the number
predicted by equation (\ref{eq:morph_gradient_soln}):
\begin{equation}
E(t) =  \frac{N(t) - \int_{0.0}^{0.5} a(x,t) \; \mbox{d} x}{\int_{0}^{\infty} a(x,t)\; \mbox{d} x}.
\label{errorEr}
\end{equation}
For the static interface (TRM) case ($I(t) \equiv 0.5$), this corresponds to
comparing the total number of molecules in the compartment region with the expected
amount. For the moving interface (ATRM) this is not the case, as the location of the
interface varies with time. However, the movement threshold $c_{max}$ is set so
that the average steady state position of the interface is at $x = 0.5$,
and therefore at steady state the position of the interface in the ATRM
simulation will be consistent with the TRM case.

\begin{figure}[htbp]
\begin{center}
  \includegraphics[width=0.5\textwidth]{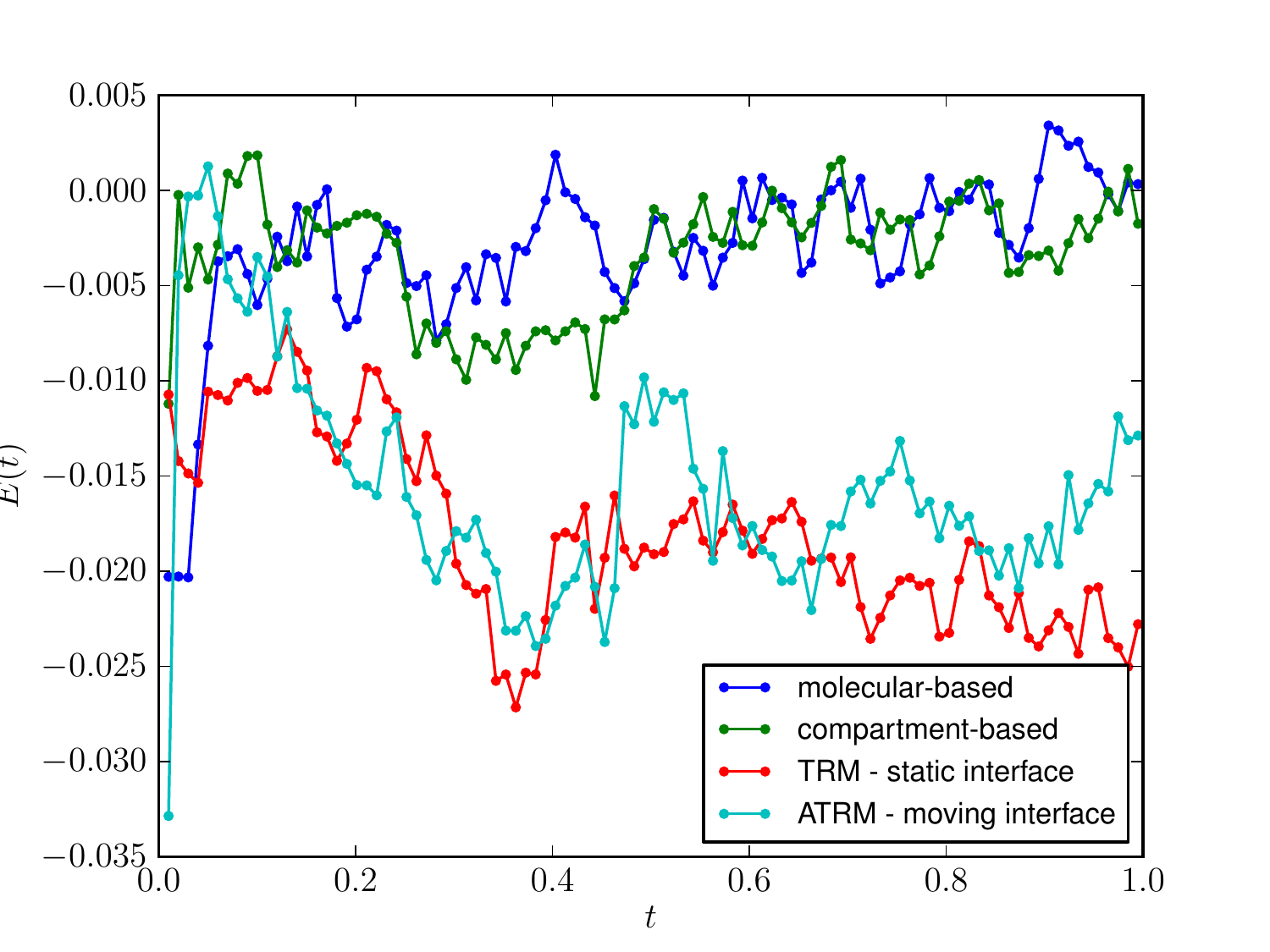}
  \caption{{\it Error $E(t)$ given by $(\ref{errorEr})$ for four different
  simulation methods, pure molecular-based and compartment-based, 
  along with the {\rm TRM} and {\rm ATRM} results.}}
  \label{fig:error_versus_time}
\end{center}
\end{figure}

Figure \ref{fig:error_versus_time} shows the error $E(t)$ versus time for the
static and moving boundary cases, along with a purely molecular-based simulation
and a compartment-based simulation. The compartment-based simulation has its
domain truncated at $x=2$ (with a reflective boundary condition). We expect
only a very small number of molecules to reach $x=2$ so this truncation will
introduce negligible error.
At very small times the error measure is dominated by the very low copy numbers
in all simulations, but after $t>0.1$ it can be seen that both TRM and ATRM simulations
lose molecules more rapidly from the $x<0.5$ region due to an overestimation of
the diffusion across the interface. The absolute value of the error $E(t)$
increases until $t\approx0.4$, where it levels out at relatively low 2\% of the total number of molecules. The
effect of the moving boundary is small for these parameters, and the net
effect of the moving boundary is to slightly decrease the diffusion of molecules
across the interface.
The next section explores further the moving interface error during the steady
state, how this varies with the simulation parameters, and therefore, how it 
may be reduced.

\subsubsection{Steady State Regime and Parameter Study}

\noindent
This section explores the effect of simulation parameters on the steady
state error for static and moving interface simulations of the morphological
gradient. Since we vary the compartment size $h$ during these parameter sweeps, the
concentration histograms are calculated using bins with a constant size of
$0.05$ along the $x$-axis (and with size $1$ along the $y$ and $z$ axis). The
contribution of each compartment $i$ to bin $j$ is scaled by the volume of $i$
that overlaps with $j$.
In order to calculate the steady state error, $E(t)$ is averaged over $40$
equally spaced times after steady state is reached at $t=5$ using 
\begin{equation}\label{eq:ssError}
\hat{E} = \sum_{i=0}^{39} E(5 + i\tau)
\end{equation}
where the spacing between each sample ($\tau=0.1$) is long enough so 
that there is no significant correlation between them.

\begin{figure*}[htbp]
\begin{center}
\centerline{
\hskip 1.2cm
\includegraphics[height=6.6cm]{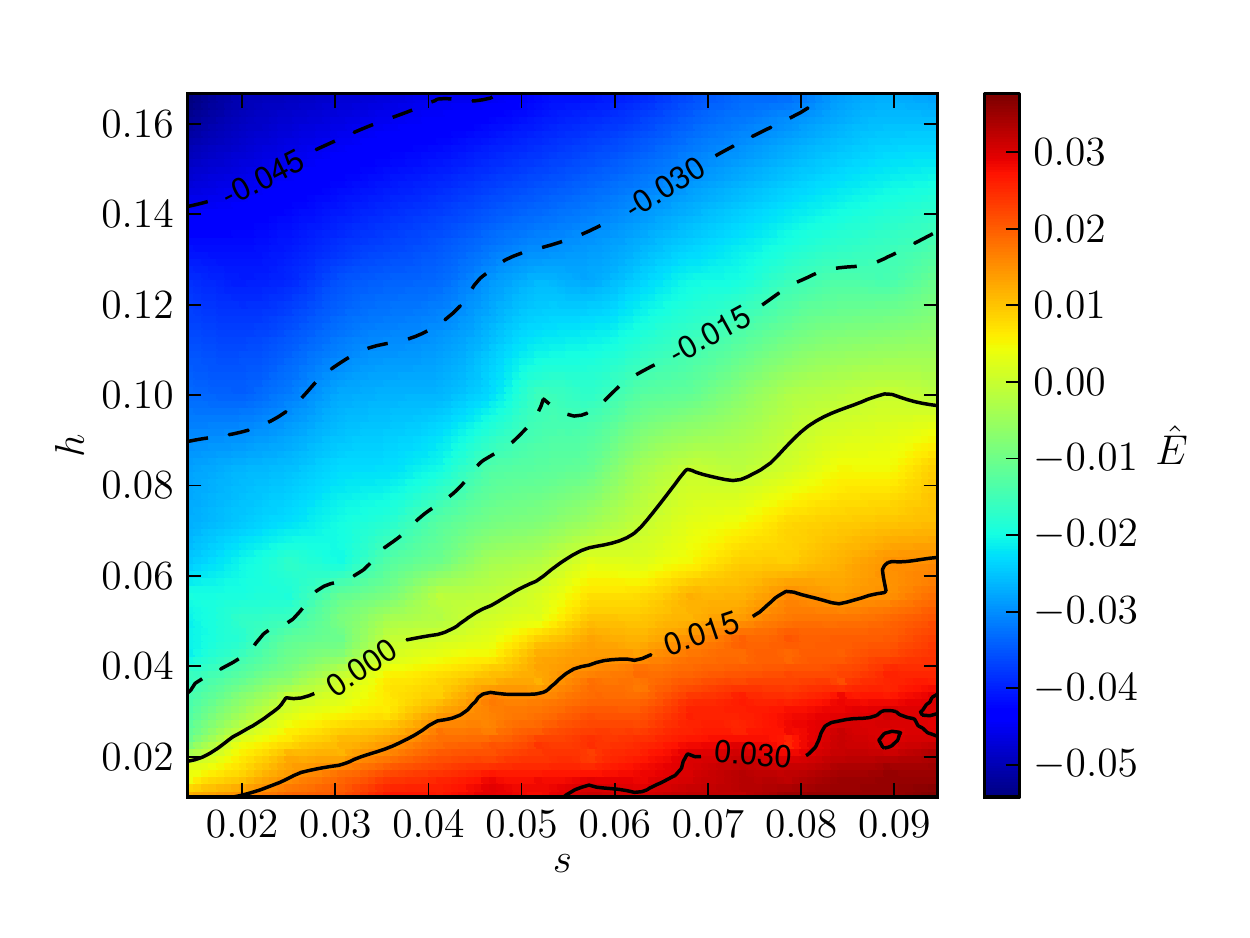}
\hskip -8.7cm
\raise 6cm \hbox{(a)}
\hskip 8.7cm
\includegraphics[height=6.6cm]{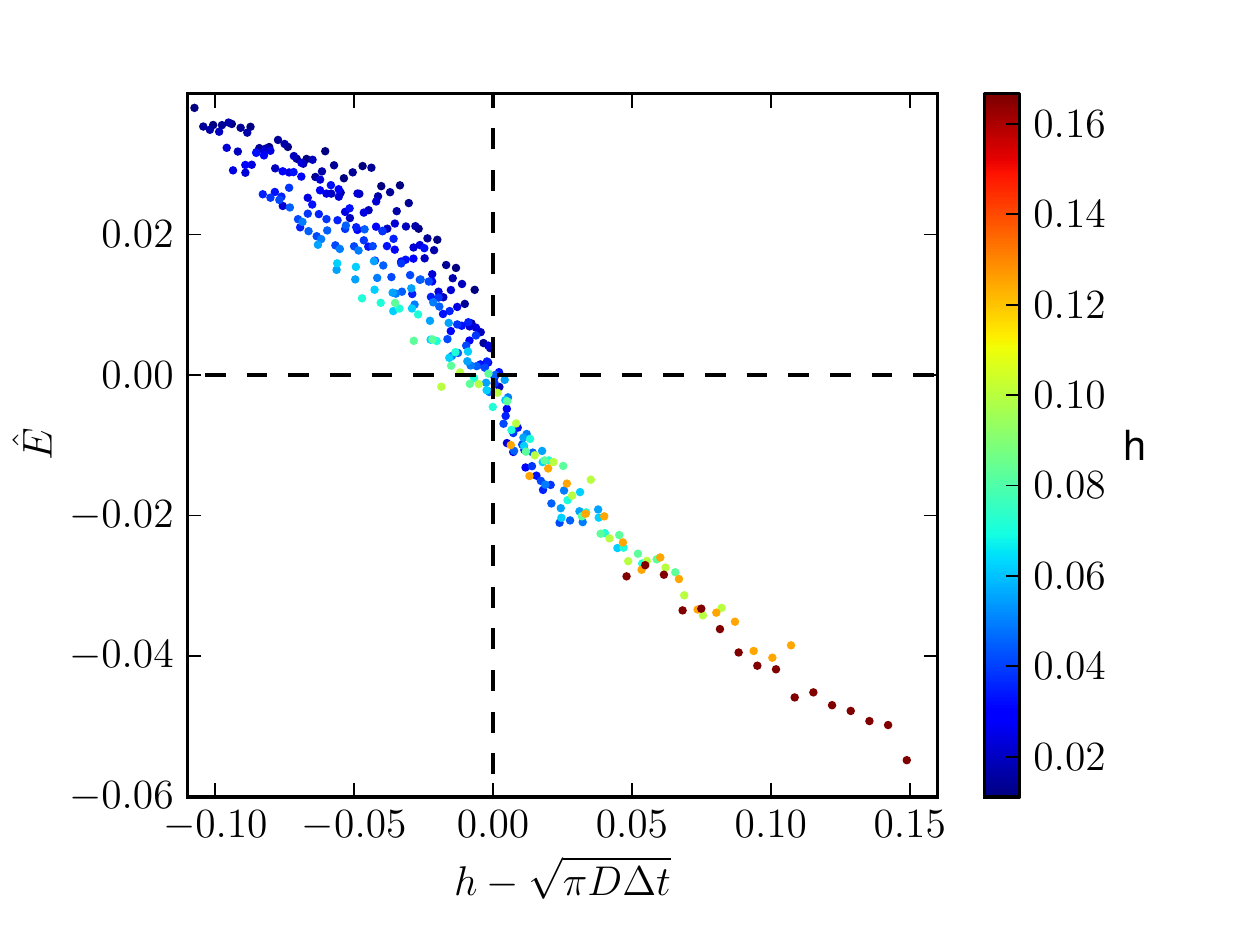}
\hskip -8.5cm
\raise 6cm \hbox{(b)}
\hskip 8.5cm
}  
\caption{(a) {\it Colorplot of the {\rm TRM} error $\hat{E}$ given by
  equation $(\ref{eq:ssError})$ for steady state morphological gradient versus 
  compartment size $h$ and particle average step size $s=\sqrt{2 D \Delta t}$;}
  (b) {\it same data as a scatter plot of $\hat{E}$ versus $h-\sqrt{\pi D \Delta
  t}$. Each point is coloured by $h$.}}
  \label{fig:param_study_si}
\end{center}
\end{figure*}

Figure \ref{fig:param_study_si}(a) shows the average steady state error
$\hat{E}$ versus compartment size $h$ and average particle step size $s=\sqrt{2
D \Delta t}$.
The contour lines of constant $\hat{E}$ generally follow a linear relationship
between $h$ and $s$, and the error is minimized near $h=s$. This is consistent
with the convergence study described in \cite{Flegg:2013:CMC}, which
found that the TRM error for a static interface was minimized when $h=\sqrt{\pi
D \Delta t}$.
Figure \ref{fig:param_study_si}(b) shows the same data in a scatter plot of
$\hat{E}$ versus $h-\sqrt{\pi D \Delta t}$, with each point is coloured by $h$.
For a given $h$ the error is linear with $h-\sqrt{\pi D \Delta t}$ around
the point $h=\sqrt{\pi D \Delta t}$, with a slope
that varies with $h$. The change in slope with $h$ is due to the $\mathcal{O}(h^2)$
diffusion error in the compartment region, and is not seen in Flegg et al
\cite{Flegg:2013:CMC} since $h$ is only refined near the interface. In
our simulations $h$ is refined over the entire compartment region and the
$\mathcal{O}(h^2)$ diffusion error becomes significant.

\begin{figure*}[htbp]
\begin{center}
\centerline{
\hskip 1.2cm
\includegraphics[height=6.6cm]{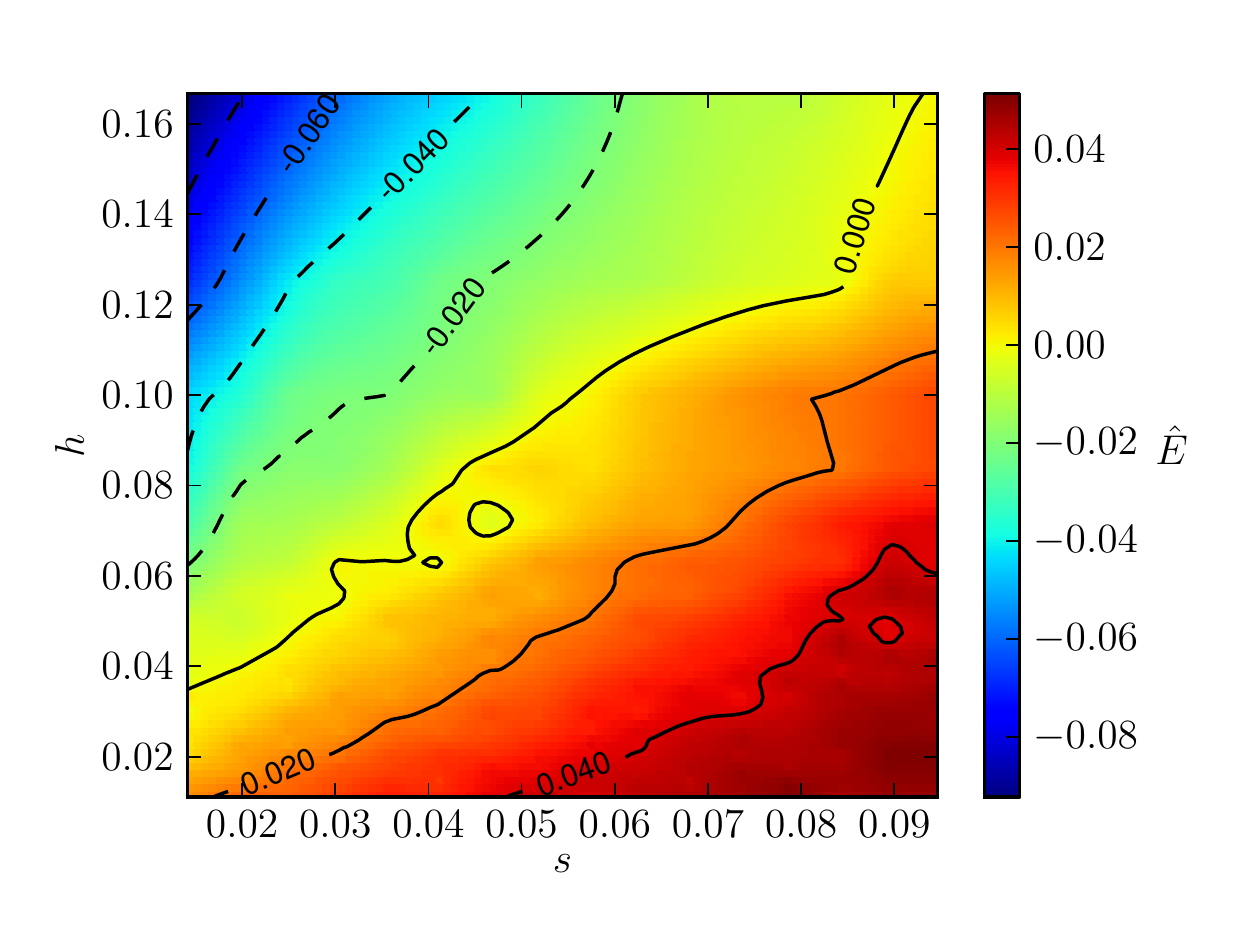}
\hskip -8.7cm
\raise 6cm \hbox{(a)}
\hskip 8.7cm
\includegraphics[height=6.6cm]{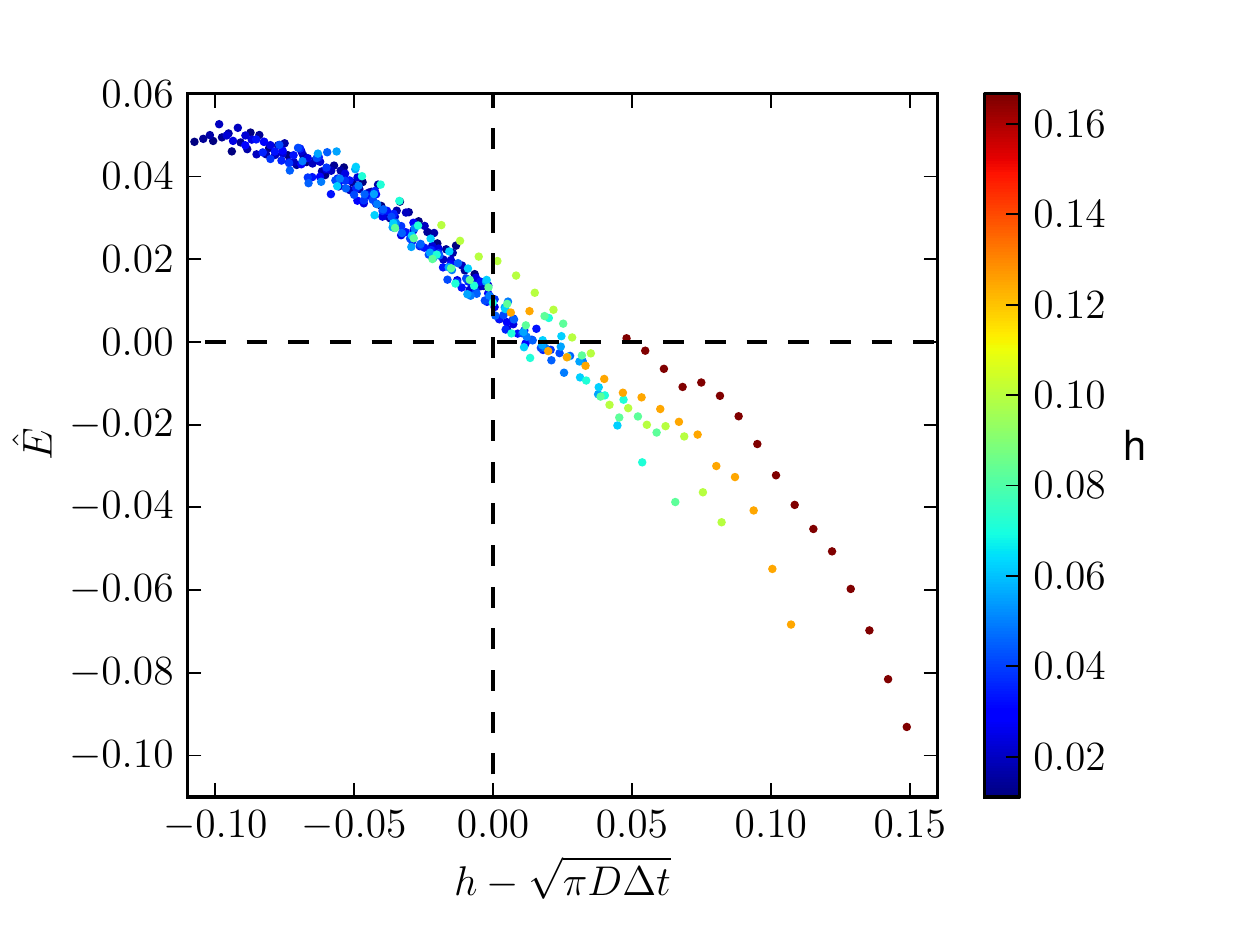}
\hskip -8.7cm
\raise 6cm \hbox{(b)}
\hskip 8.7cm
}  
  \caption{(a) {\it Colorplot of the {\rm ATRM} error $\hat{E}$ given by
  equation $(\ref{eq:ssError})$  for steady
  state morphological gradient versus compartment size $h$ and particle average step
  size $s=\sqrt{2 D \Delta t}$;}
  (b) {\it same data as a scatter plot of $\hat{E}$ versus $h-\sqrt{\pi D \Delta
  t}$. Each point is coloured by $h$.}}
  \label{fig:param_study_mi}
\end{center}
\end{figure*}

Figure \ref{fig:param_study_mi} shows the steady state error $\hat{E}$ for the
ATRM simulation with moving interface. The most obvious change in $\hat{E}$ with 
the moving interface is the shifting of the plots towards positive $\hat{E}$. 
That is, the flux of molecules across the interface towards the molecular region is
(slightly) reduced.
For intermediate and small values of $h$ this reduction is small (1-2\%), but
for larger $h$ the shifts become more pronounced due to much larger jumps that
the interface makes. The scatter plot in Figure \ref{fig:param_study_mi}(right)
also shows a strong non-linear reduction in $\hat{E}$ for $h \gg \sqrt{\pi D
\Delta t}$.

In summary, the effect of the moving interface on the error associated with the
TRM is minimal, and generally in the region of 1-2\% of the expected
molecule concentration. This increases for larger $h$ due to the larger step
size of the interface, but remains relatively small (less than 3\%) unless $h \gg
\sqrt{\pi D \Delta t}$, when it starts to diverge.

\begin{figure}[htbp]
\begin{center}
  \includegraphics[width=0.5\textwidth]{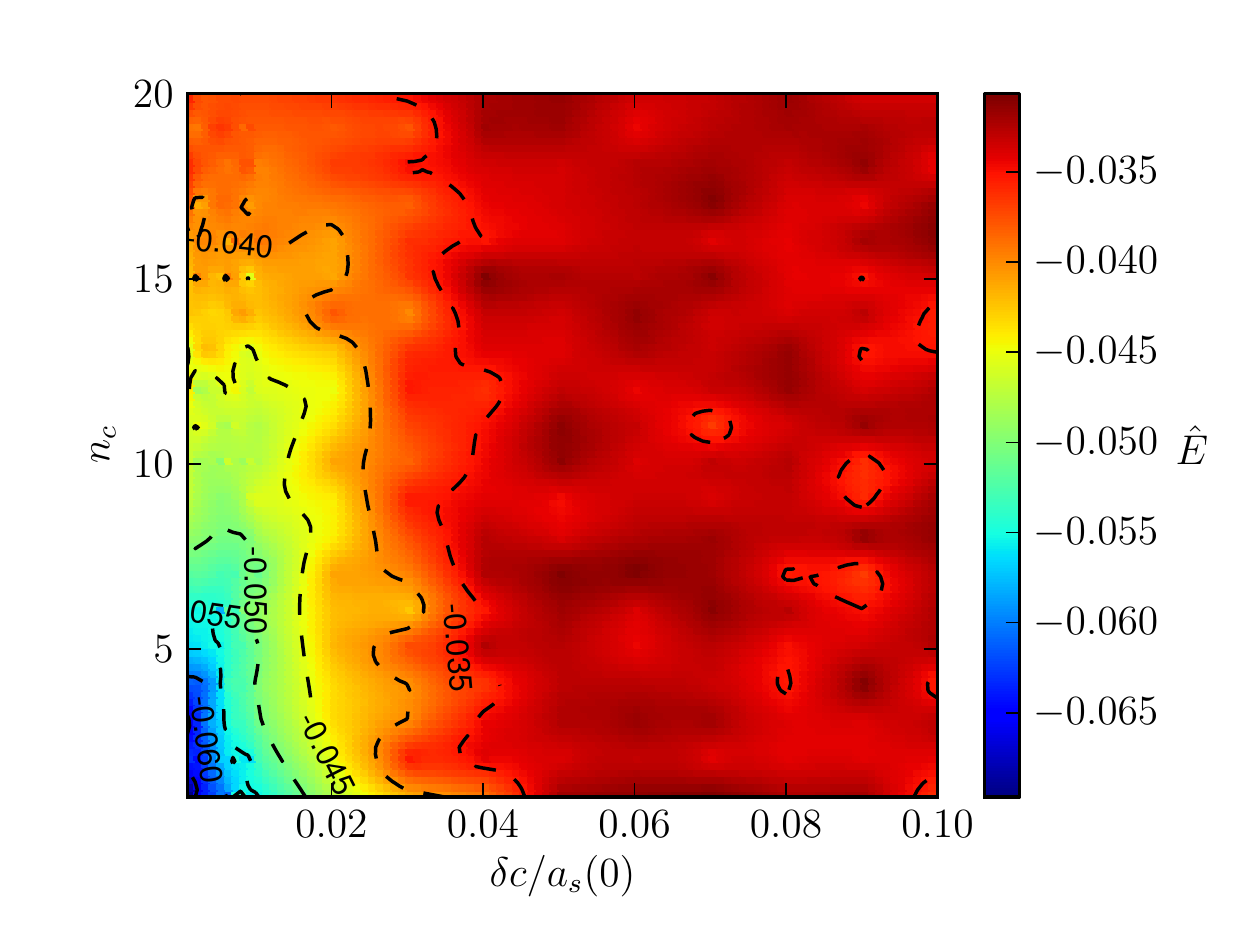}
  \caption{{\it Colorplot of the {\rm ATRM} error $\hat{E}$ given by equation
  $(\ref{eq:ssError})$ for steady state morphological gradient versus threshold
  separation $\delta c$ and $n_c$ of the {\rm ATRM}.}}
  \label{fig:param_study_mi2}
\end{center}
\end{figure}

Figure \ref{fig:param_study_mi2} shows the same steady state error $\hat{E}$ for
the moving interface versus the ATRM parameters used to specify the movement criteria.
These are $\delta c$, the separation between the upper and lower
thresholds for movement, and $n_c$, the number of timesteps between checks of 
the movement
criteria. For these simulations the resolution parameters are kept constant at
$h=0.1$ and $s=\sqrt{2 D \Delta t}=0.014$. The results of this parameter sweep 
show that $\delta c$
has the greatest effect on the error. The error decreases as $\delta c$ is
increased, and for these parameters is minimized for $\delta c$ greater than 4\%
of the maximum steady state concentration $a_s(0)$. The error decreases more
slowly for increasing $n_c$, and we also note that the increase in $n_c$ also restricts the maximum speed of the moving interface. It
is therefore clear that increasing $\delta c$ is the optimal method to reduce
the error associated with the moving interface.

\subsection{Fisher Wave}\label{sec:fisher_wave}

\noindent
The Fisher equation \cite{Fisher:1937:WAA} is the prototype
model for the spread of a biological species and describes the diffusive spread
of a species along with a logistic growth term
\begin{equation}\label{eq:fisher_eq}
\frac{\partial u}{\partial t} = D \triangle u + k_1 u - k_2 u^2
\end{equation}
Given the phase space $(u,\frac{\partial u}{\partial t})$, the Fisher equation has
unstable stationary point at $(0,0)$ leading to a stable node at $(k_1/k_2,0)$. 
It admits travelling wave solutions that transition from the
unstable to the stable stationary point, which move with a wave speed $c\ge2$.
The wave will move with its minimum wave speed $c=2$ as long as the initial
condition $u(x,0)$ is zero outside a finite domain
\cite{Murray:2002:MB}.

A single-species stochastic reaction-diffusion system matching the above PDE
model can be constructed. Consider the evolution of a single species $A$ 
which undergoes diffusion and a reversible reaction 
(\ref{reversiblereaction}).
Assuming a large number of molecules, the mean-field concentration of species
$A$ will approach equation (\ref{eq:fisher_eq}). 
However, for low molecule copy numbers, stochastic effects can play an increasing
role in the dynamics of the system.
Numerous lattice-based models have shown that the stochastic fluctuations in the
number of $A$ molecules act to reduce the wave speed by a term $c^*$
proportional to $\log^{-2}N_0$, where $N_0$ is the average number of $A$ molecules
in each lattice site behind the wavefront
\cite{Panja:2004:EFP,vanSaarloos:2003:FPU,Brunet:2001:EMN}. This result, however, is
not immediately applicable to molecular-based models, since $N_0$ is inversely
proportional to the volume of each lattice site and thus is
determined by the lattice itself. While it would be useful to establish a
similar scaling law for molecular-based methods, the
computational requirements of such methods scale quickly with increasing
molecule numbers and it is therefore difficult to run the large simulations that
are needed to approach the corresponding mean-field model.

The problem of running a stochastic travelling wave simulation with high
molecule numbers is ideal for the ATRM. Setting the location of interface $I(t)$
directly behind the wavefront means that the high
concentration region behind the wave is modelled by the compartment-based
method, while the wavefront itself and the low concentration region in front of
the wave is modelled by the molecular-based method. Therefore the wave dynamics are
captured entirely by the molecular-based method, while the total number of
discrete molecules simulated is small and restricted only to
those that can affect the wave propagation.

\begin{figure}[htbp]
\begin{center}
  \includegraphics[width=0.5\textwidth]{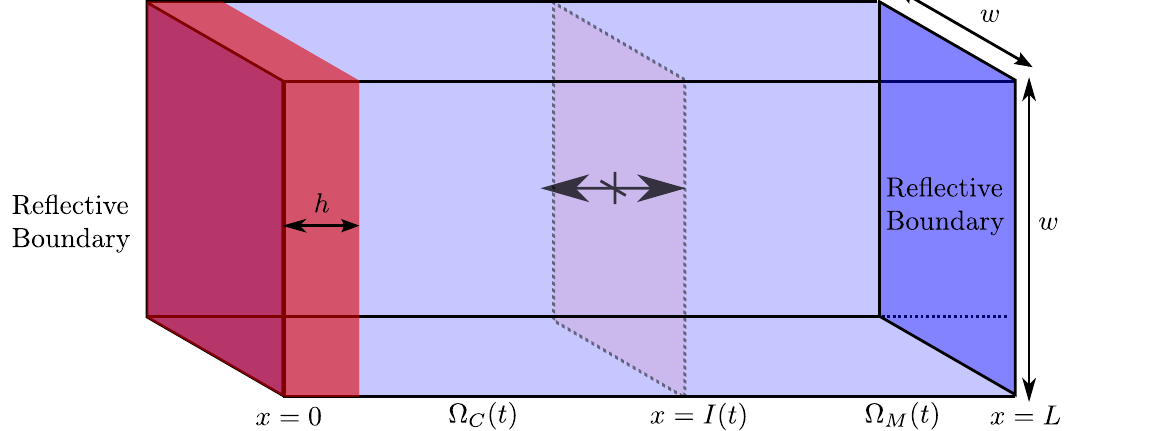}
  \caption{{\it Fisher wave simulation domain. Upper and lower $x$-axis boundaries
  are reflective. All other boundaries are periodic. The position of the
  (moving) interface between the molecular-based $\Omega_M$ and
  compartment-based $\Omega_C$ regions is $x=I(t)$. At $t=0$ the interface is
  located at $x=I(0)=h$, and the region $x<h$ is filled with $h \, w^2 \, k_1/k_2$
  particles that are placed randomly within the region. The domain length is
  set to $L$ and the height and depth of the domain are set to
  $w$. }}
  \label{fig:fisher_sim_domain}
\end{center}
\end{figure}

Figure \ref{fig:fisher_sim_domain} shows the simulation domain $\Omega = (0,L)
\times (0,w) \times (0,w)$. The initial conditions of the simulation are a
random and homogeneous distribution of $A$ molecules with concentration $k_1/k_2$ over the volume defined by $0 \leq x \leq
h$ and $0 \leq y,z \leq w$. The domain in the $y$ and $z$ directions is periodic
with length $w$, therefore the travelling wave will propagate as a
one-dimensional wave in the positive $x$ direction. The domain width is scaled
by the expected concentration behind the wavefront $w = \sqrt{800 \, k_1/k_2}$
in order to keep the total number of molecules constant with a varying reaction
ratio $k_1/k_2$. The lower and upper $x$ boundaries are both reflective. The
interface between $\Omega_M$ and $\Omega_C$ is a plane with normal parallel to
the $x$-axis and it moves with a step-size $h$. The parameters of the
Fisher wave simulation are given in Table \ref{tab:fisherwave}.
Three different stochastic simulations were run using (a) a purely
compartment-based method, (b) a molecular-based method and (c) the ATRM method.
A snapshot of each simulation taken at $t=40$ was shown in the introduction in
Figure \ref{fig:default_fisher_sim}.
\begin{table}
\begin{center}
\begin{tabular}{|c | c|}
\hline
Parameter & Value \\
\hline
$D$ & 1 \\
$k_1$ & 1 \\
$k_2$ & 1 \\
\hline
$\rho$ & 0.5 \\
$\alpha$ & 0.6 \\
$P_{\Delta t}$ & $3.7 \times 10^{-3}$ \\
\hline
$h$ & 2.5 \\
$\Delta t$ & $10^{-3}$ \\
$L$ & 100 \\
$w$ & 28.3 \\
\hline
$c_{max}$ & $0.95\ k_1/k_2 = 0.95$ \\
$\delta c$ & $0.55\ k_1/k_2 = 0.55$ \\
$n_c$ & 10 \\
\hline
\end{tabular}
\caption{{\it Table of parameters for the Fisher wave simulation,
used in Figure  $\ref{fig:default_fisher_sim}$.
The first three parameters are the parameters of the biological model
($D$, $k_1$ and $k_2$). Using $(\ref{calcro})$, they were transformed
to binding and unbinding radii $\rho$ and $\alpha \rho$.
Parameter $h$ is the compartment size in $\Omega_C$
and parameter $\Delta t$ is the time step in $\Omega_M$. The last three 
parameters $c_{max}$, $\delta c$ and $n_c$ are the parameters of the {\rm ATRM}.}}
\label{tab:fisherwave}
\end{center}
\end{table}

Our goal here is to ensure that the more efficient ATRM simulation matches the
results obtained by the molecular-based method, and this is indeed the case.
In Figure \ref{fig:default_fisher_sim}, both the ATRM (bottom panel) and 
the molecular-based (middle panel) simulation are very similar
in terms of both the wavefront shape and propagation speed. However, clear
differences can be seen in these wave speeds and those of the mean-field model
and compartment-based simulation. These differences in wave speed and the effect
of the parameters $k_1/k_2$ and $h$ are explored further on in this section.

\begin{figure}[htbp]
\begin{center}
  \includegraphics[width=0.5\textwidth]{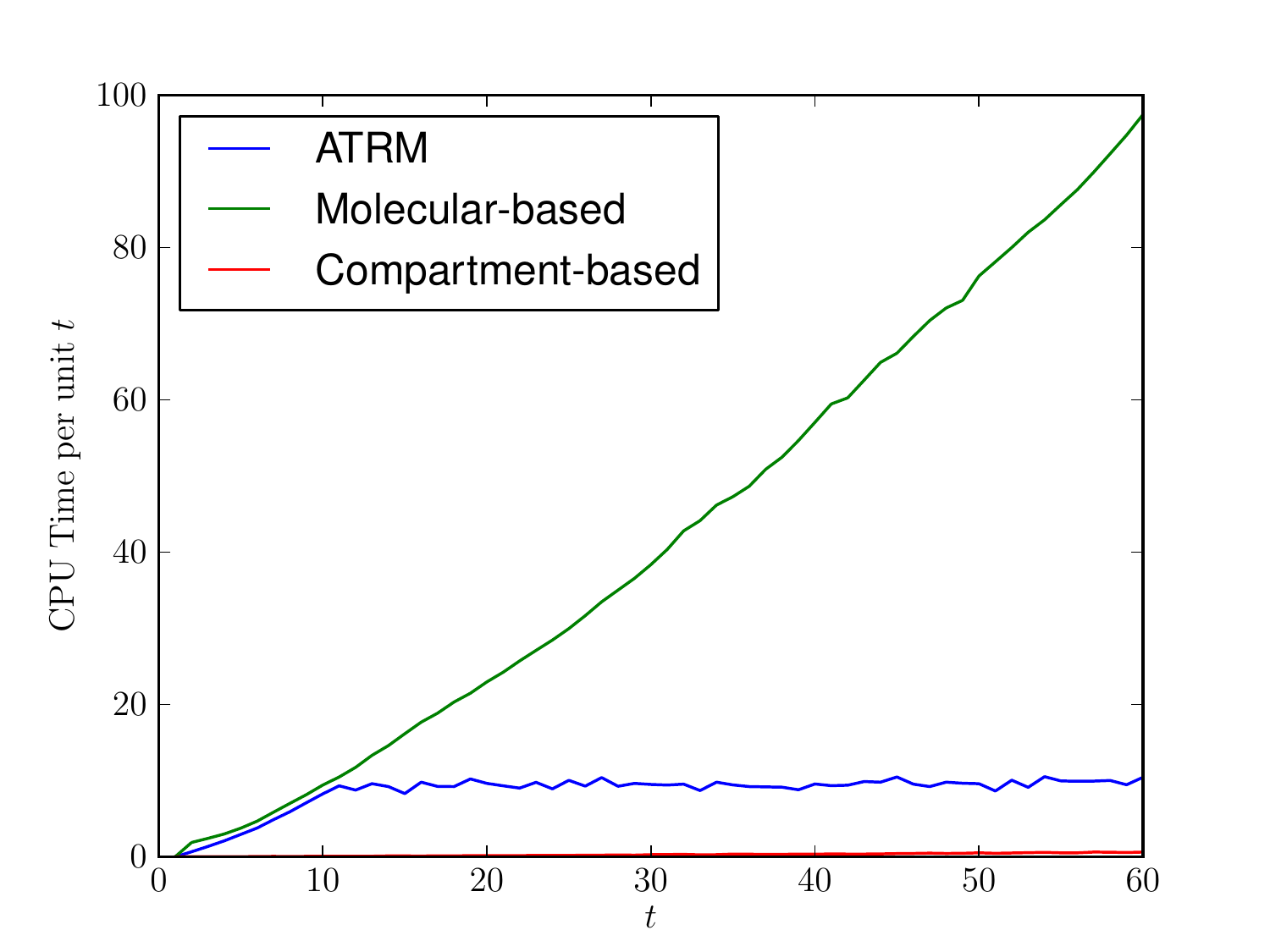}
  \caption{{\it Comparison of computational time for each of the simulation
  methods. The plots show the CPU time taken to simulate 1 second of each model
  as a function of time $t$. }}
  \label{fig:fisher_times}
\end{center}
\end{figure}

To demonstrate the efficiency gained by using the ATRM method, Figure
\ref{fig:fisher_times} shows a comparison of the time taken to run each of the
three different simulation methods. The plots show the CPU time taken to
complete 1\% of simulation time versus the total simulation percentage
performed. The purely compartment-based method (red line) is clearly the
fastest, and its plot can barely be seen at the bottom of Figure
\ref{fig:fisher_times}. The purely
molecular-based simulation is the slowest. 
The ATRM simulation initially follows the molecular-based simulation, until the
interface starts moving to follow the travelling wave at about $t=10$.
After this point there is a constant number of discrete molecules in the
simulation (those in the wavefront itself) and therefore the simulation CPU time
remains roughly constant.

\begin{figure}[htbp]
\begin{center}
  \includegraphics[width=0.5\textwidth]{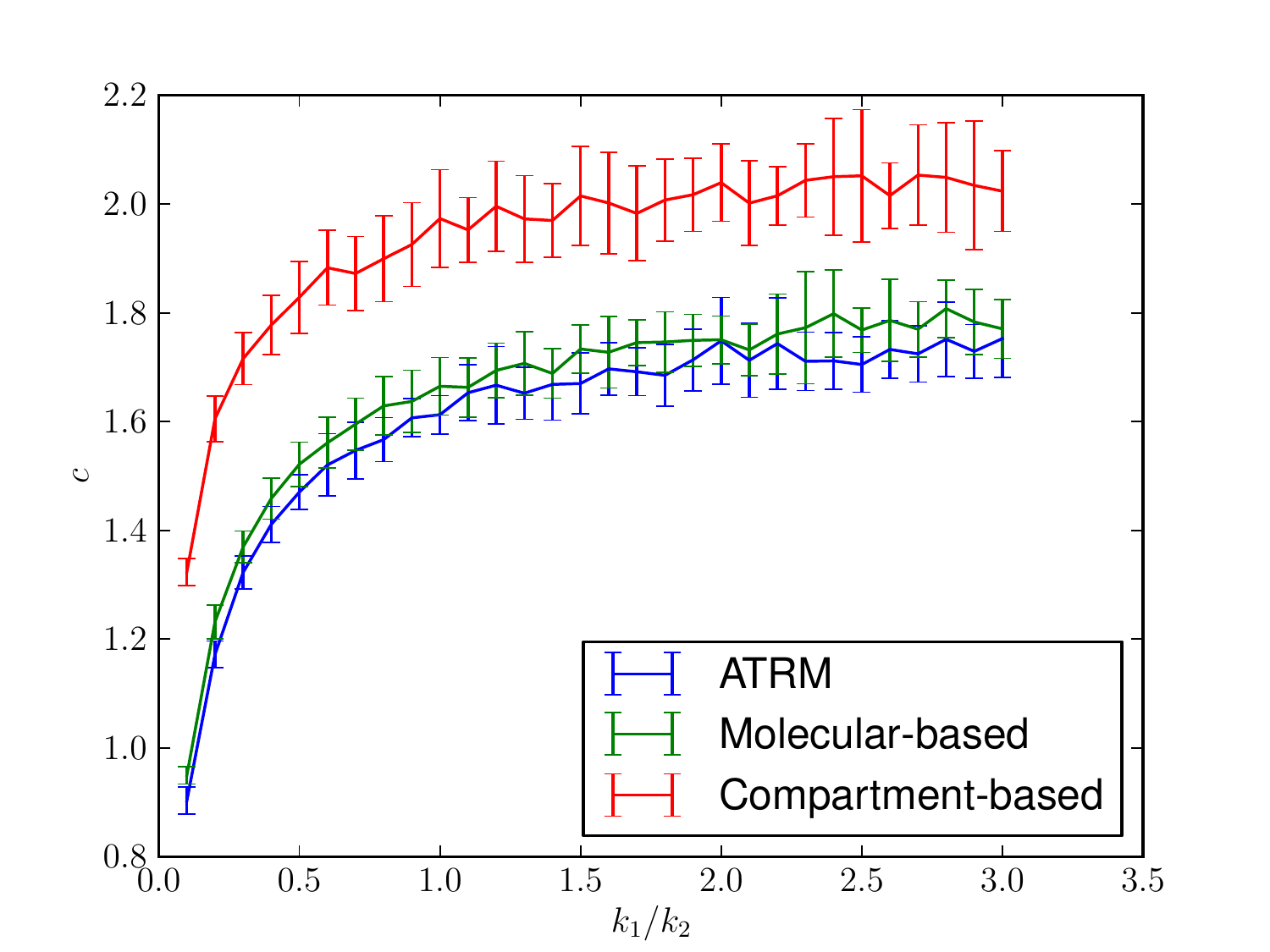}
  \caption{{\it Wave speed versus $k_1/k_2$ for the three simulation
  methods. The wave speed is estimated using $(\ref{estwavespeed})$ from 20 different
  simulations, the average of these gives the solid line and the
  error bars show one standard deviation.}}
  \label{fig:fisher_speed_versus_n}
\end{center}
\end{figure}

For a travelling wave simulation the important output measurement is normally
the wave speed. Any stochastic simulation method must be able to
accurately reproduce the speed of the wave and correctly capture any stochastic
effects. Therefore we have measured the simulated wave speed versus the  
reaction ratio $k_1/k_2$ and the compartment size $h$. 

The reaction ratio $k_1/k_2$ determines the saturation concentration of the wave
(the concentration behind the wavefront) and therefore the number of molecules
in the wavefront. Increasing this parameter increases the number of molecules in
any given volume and therefore we would expect the wave speed to approach the
mean-field wave speed $c=2$ as $k_1/k_2$ increases for the compartment-based
model. In the case of the molecular-based models, the mean-field PDE description
is often justified under special circumstances (e.g. for systems with uniformly
distributed reactants) and the convergence of travelling speeds to the
mean-field model is not obvious.

Figure \ref{fig:fisher_speed_versus_n} shows the measured wave speed $c$ versus
$k_1/k_2$ for the three different simulation methods. The wave speed is measured
as follows. Given the total number of molecules at a given time during the
simulation $N_{tot}(t)$, we can obtain the estimate of the wave speed
$c$ as the appropriately rescaled rate of change of $N_{tot}$ between two
times $t_1$ and $t_2$:
\begin{equation}
c = \frac{(N_{tot}(t_2)-N_{tot}(t_1))}{(t_2-t_1)}\frac{k_2}{k_1 \, w^2}. 
\label{estwavespeed}
\end{equation}
For each parameter value, we ran 20 Fisher wave simulations and calculate the
wave speed using using $t_1 = 10$ and $t_2 = 30$. The mean wave speed is plotted
in Figure \ref{fig:fisher_speed_versus_n} as a solid line, while the error bars
show one standard deviation.

\begin{figure}[htbp]
\begin{center}
  \includegraphics[width=0.5\textwidth]{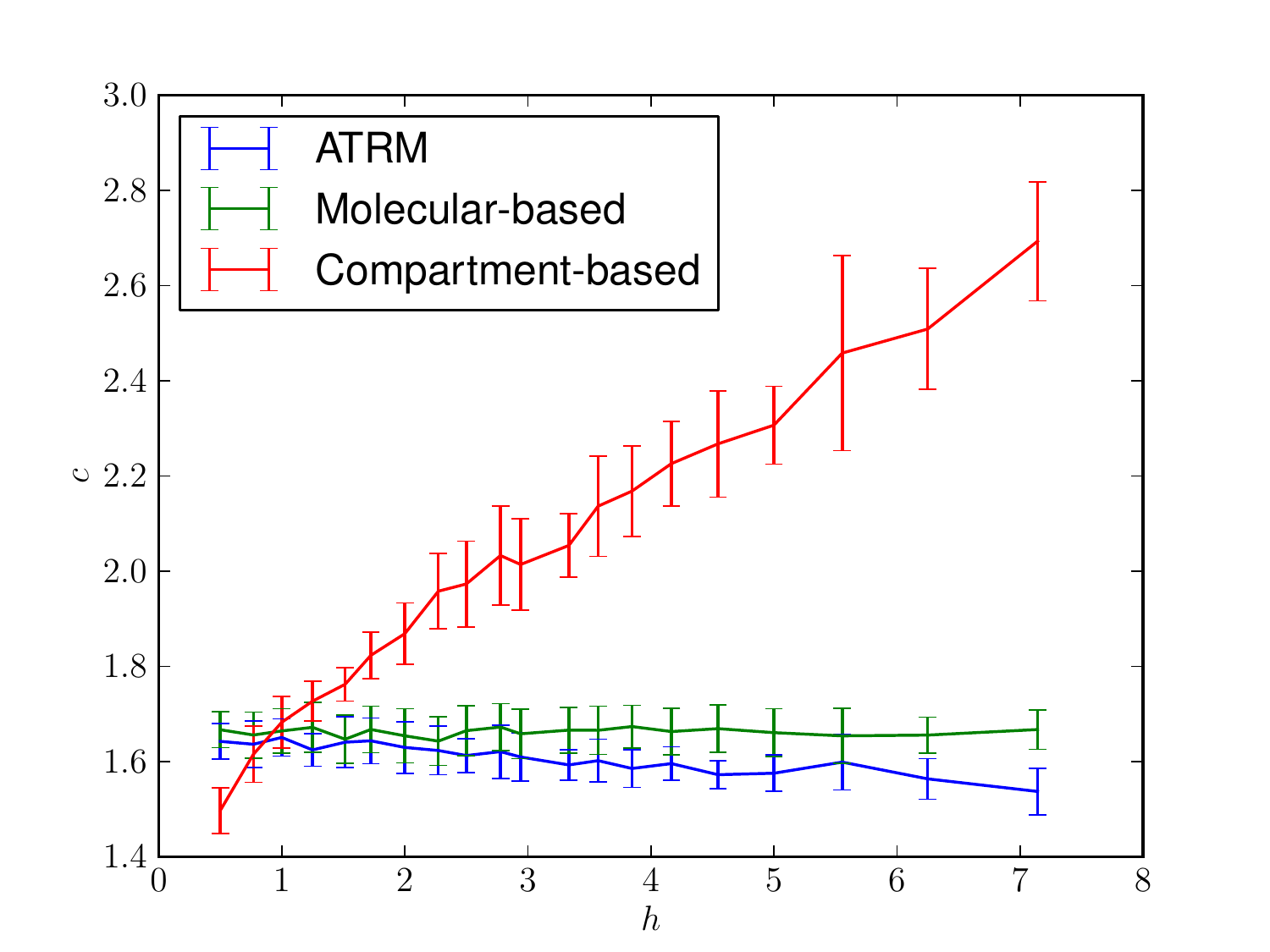}
  \caption{{\it Wave speed versus compartment size $h$ for the three simulation
  methods. The wave speed is estimated using $(\ref{estwavespeed})$ from 20 different
  simulations, the average of these gives the solid line and the
  error bars show one standard deviation.}}
  \label{fig:fisher_speed_versus_h}
\end{center}
\end{figure}

The results show that the ATRM method with moving interface produces identical
results to the purely molecular-based simulation for all values of $k_1/k_2$, within
the range of stochastic fluctuations for the wave speed. As stated earlier, our
goal is to match the results of the molecular-based method, which is achieved
here. Note that the compartment-based method, while producing a similar scaling
with $k_1/k_2$, gives a consistently higher wave speed than either of the other
methods. This change in wave speed for the compartment-based method was found to
vary with the compartment size $h$, and this is shown later on in Figure
\ref{fig:fisher_speed_versus_h}. However, while the ATRM simulation uses the
compartment-based method for the domain behind the wavefront $\Omega_C$, the
wave front is situated entirely in the molecular-based domain $\Omega_M$ and
thus the motion of the simulated wave is determined only by the molecular-based
method. The diffusion error introduced by the ATRM interface is very small and
has no effect on the simulation. Due to the position of the interface behind the
wave front, the local concentration gradient is zero at the interface which
results in a negligible ATRM diffusion error.

Figure \ref{fig:fisher_speed_versus_h} shows the measured wave speeds $c$ versus
the compartment size $h$. As previously stated, for the compartment-based method
the wave speeds show a clear dependence on $h$. This result is not surprising,
given that previous lattice-based simulations
\cite{Panja:2004:EFP,vanSaarloos:2003:FPU,Brunet:2001:EMN} showed a $\log^{-2}N_0$
scaling for $c$, where $N_0$ is the average number of molecules per lattice site
and is therefore determined by the lattice spacing. In addition, the diffusion
error in the compartment-based method is of order $\mathcal{O}(h^2)$, and for
the Fisher wave this has the effect of increasing the wave speed enough that it
becomes larger than the mean-field speed ($c=2$) for $h \ge 3$. However, neither
of these effects apply to the molecular-based method, which does not have either a
background lattice nor a set of compartments. The ATRM method, as desired, matches the molecular-based
method perfectly and displays a constant wave speed versus $h$.

\section{Discussion}
\label{secdiscussion}

\noindent
We extended the TRM to consider a moving interface between the domains that
can respond dynamically to the simulation variables (local concentration).
This can be considered as an adaptive domain decomposition method, which
motivates the addition of the word ``Adaptive'' to the TRM.
The ATRM is a multiscale method aimed at coupling compartment and
molecular-based stochastic reaction-diffusion simulations with a moving
interface. In this paper, we applied it to two different problems, 
a morphological gradient problem with a steady-state solution and 
a Fisher travelling wave where the movement of the interface 
is determined by the wave motion.

The error associated with the moving interface for the steady-state
morphological gradient case was investigated using parameter sweeps.
It was found that effect of the moving boundary was minimal and that the
additional error (over a static interface) was generally in the region of 1-2\%.
This error increases with $h$, the compartment size, and further increases for $h$ much
larger than $\sqrt{\pi D \Delta t}$, where $D$ is the diffusion constant and
$\Delta t$ is the molecular-based timestep. The interface error was insensitive
to the choice of $n_c$ (minimum number of timesteps between interface movement),
but it was found that $\delta c$ (the separation between the upper and lower
concentration thresholds) needed to be greater than 4\% of the maximum molecular
concentration for the error to be minimised.

The second test problem clearly showed the advantages of the ATRM method with a
moving interface, applying the method to a travelling wave simulation using a
single-species version of the classical Fisher equation. Here the
wavefront was simulated by the molecular-based method while the
compartment-based method was used for the high concentration region behind the
front. The ATRM simulation showed a decrease in simulation time because
the simulation time is dominated by the molecules that are in the wave front
itself, not those behind the wave front.

The wave speed $c$ was measured for varying $k_1/k_2$, the ratio of the
forward to backward reaction rates, and $h$, the compartment size. In all cases
the measured wave speed was identical for both the TRM and the molecular-based
simulations. All the methods showed an increase in wave speed with greater
$k_1/k_2$. However, the wave speed for the compartment-based method showed a
dependence on the $h$, which was not seen in either the ATRM or molecular-based results.
Therefore, the ATRM method can be seen to match the molecular-based method,
while at the same time being more computationally efficient. The error
associated with the moving interface was negligible due to the location of the
interface behind the wave front, where the concentration gradient is zero on
average. The ATRM is therefore an ideal method to study the dynamics of a
stochastic Fisher wave modelled using a molecular-based method, which would
ordinarily be impractical due to the large number of molecules needed.

Another hybrid simulation approach would be to use a mean-field, deterministic, model
for the simulation behind the wave front, which is then coupled to the
molecular-based model for the wave front. This type of model has been used previously (but not 
applied to the Fisher wave) by Alexander et. al.
\cite{Alexander:2002:ARS}, Geyer et. al. \cite{Geyer:2004:IBD} and
Wagner and Flekkoy \cite{Wagner:2004:HCF}. The disadvantage of coupling a
mean-field model to a molecular-based model is that an overlap region is
generally required in order to calculate the mass flux across the interface, and
to accurately compute variances near the interface \cite{Franz:2013:MRA}.
In contrast, using the combination of a compartment-based and molecular-based
IBMs does not necessarily require an overlap region, and retains the stochastic 
nature of the model over the entire domain. We have also found the computational 
expense of the compartment-based model to be insignificant compared with the time 
spent on the molecular-based model (see Figure \ref{fig:fisher_times}), so there 
is little
motivation to use a mean-field model instead.

\section*{Acknowledgements}
This publication arises from research funded by the John Fell Oxford University
Press (OUP) Research Fund. The research leading to these results has received
funding from the European Research Council under the \emph{European Community’s}
Seventh Framework Programme \emph{(FP7/2007-2013)} / ERC \emph{grant agreement}
No. 239870. Radek Erban would also like to thank Brasenose College, University
of Oxford, for a Nicholas Kurti Junior Fellowship; the Royal Society for a University
Research Fellowship; and the Leverhulme Trust for a Philip Leverhulme Prize.


\begin{thebibliography}{38}%
\makeatletter
\providecommand \@ifxundefined [1]{%
 \@ifx{#1\undefined}
}%
\providecommand \@ifnum [1]{%
 \ifnum #1\expandafter \@firstoftwo
 \else \expandafter \@secondoftwo
 \fi
}%
\providecommand \@ifx [1]{%
 \ifx #1\expandafter \@firstoftwo
 \else \expandafter \@secondoftwo
 \fi
}%
\providecommand \natexlab [1]{#1}%
\providecommand \enquote  [1]{``#1''}%
\providecommand \bibnamefont  [1]{#1}%
\providecommand \bibfnamefont [1]{#1}%
\providecommand \citenamefont [1]{#1}%
\providecommand \href@noop [0]{\@secondoftwo}%
\providecommand \href [0]{\begingroup \@sanitize@url \@href}%
\providecommand \@href[1]{\@@startlink{#1}\@@href}%
\providecommand \@@href[1]{\endgroup#1\@@endlink}%
\providecommand \@sanitize@url [0]{\catcode `\\12\catcode `\$12\catcode
  `\&12\catcode `\#12\catcode `\^12\catcode `\_12\catcode `\%12\relax}%
\providecommand \@@startlink[1]{}%
\providecommand \@@endlink[0]{}%
\providecommand \url  [0]{\begingroup\@sanitize@url \@url }%
\providecommand \@url [1]{\endgroup\@href {#1}{\urlprefix }}%
\providecommand \urlprefix  [0]{URL }%
\providecommand \Eprint [0]{\href }%
\providecommand \doibase [0]{http://dx.doi.org/}%
\providecommand \selectlanguage [0]{\@gobble}%
\providecommand \bibinfo  [0]{\@secondoftwo}%
\providecommand \bibfield  [0]{\@secondoftwo}%
\providecommand \translation [1]{[#1]}%
\providecommand \BibitemOpen [0]{}%
\providecommand \bibitemStop [0]{}%
\providecommand \bibitemNoStop [0]{.\EOS\space}%
\providecommand \EOS [0]{\spacefactor3000\relax}%
\providecommand %\BibitemShut  [1]{\csname bibitem#1\endcsname}%
\let\auto@bib@innerbib\@empty
%</preamble>
\bibitem [{\citenamefont {Murray}(2002)}]{Murray:2002:MB}%
  \BibitemOpen
  \bibfield  {author} {\bibinfo {author} {\bibfnamefont {J.}~\bibnamefont
  {Murray}},\ }\href@noop {} {\emph {\bibinfo {title} {{M}athematical
  {B}iology}}}\ (\bibinfo  {publisher} {Springer Verlag},\ \bibinfo {year}
  {2002})%\BibitemShut {NoStop}%
\bibitem [{\citenamefont {Black}\ and\ \citenamefont
  {McKane}(2012)}]{Black:2012:SFE}%
  \BibitemOpen
  \bibfield  {author} {\bibinfo {author} {\bibfnamefont {A.}~\bibnamefont
  {Black}}\ and\ \bibinfo {author} {\bibfnamefont {A.}~\bibnamefont {McKane}},\
  }\href@noop {} {\bibfield  {journal} {\bibinfo  {journal} {Trends in Ecology
  and Evolution}\ }\textbf {\bibinfo {volume} {27}},\ \bibinfo {pages} {337}
  (\bibinfo {year} {2012})}%\BibitemShut {NoStop}%
\bibitem [{\citenamefont {Gillespie}, \citenamefont {Hellander},\ and\
  \citenamefont {Petzold}(2013)}]{Gillespie:2013:PSA}%
  \BibitemOpen
  \bibfield  {author} {\bibinfo {author} {\bibfnamefont {D.}~\bibnamefont
  {Gillespie}}, \bibinfo {author} {\bibfnamefont {A.}~\bibnamefont
  {Hellander}}, \ and\ \bibinfo {author} {\bibfnamefont {L.}~\bibnamefont
  {Petzold}},\ }\href@noop {} {\bibfield  {journal} {\bibinfo  {journal}
  {Journal of Chemical Physics}\ }\textbf {\bibinfo {volume} {138}},\ \bibinfo
  {pages} {170901} (\bibinfo {year} {2013})}%\BibitemShut {NoStop}%
\bibitem [{\citenamefont {Andrews}\ and\ \citenamefont
  {Bray}(2004)}]{Andrews:2004:SSC}%
  \BibitemOpen
  \bibfield  {author} {\bibinfo {author} {\bibfnamefont {S.}~\bibnamefont
  {Andrews}}\ and\ \bibinfo {author} {\bibfnamefont {D.}~\bibnamefont {Bray}},\
  }\href@noop {} {\bibfield  {journal} {\bibinfo  {journal} {Physical Biology}\
  }\textbf {\bibinfo {volume} {1}},\ \bibinfo {pages} {137} (\bibinfo {year}
  {2004})}%\BibitemShut {NoStop}%
\bibitem [{\citenamefont {Lipkova}\ \emph {et~al.}(2011)\citenamefont
  {Lipkova}, \citenamefont {Zygalakis}, \citenamefont {Chapman},\ and\
  \citenamefont {Erban}}]{Lipkova:2011:ABD}%
  \BibitemOpen
  \bibfield  {author} {\bibinfo {author} {\bibfnamefont {J.}~\bibnamefont
  {Lipkova}}, \bibinfo {author} {\bibfnamefont {K.}~\bibnamefont {Zygalakis}},
  \bibinfo {author} {\bibfnamefont {J.}~\bibnamefont {Chapman}}, \ and\
  \bibinfo {author} {\bibfnamefont {R.}~\bibnamefont {Erban}},\ }\href@noop {}
  {\bibfield  {journal} {\bibinfo  {journal} {SIAM Journal on Applied
  Mathematics}\ }\textbf {\bibinfo {volume} {71}},\ \bibinfo {pages} {714}
  (\bibinfo {year} {2011})}%\BibitemShut {NoStop}%
\bibitem [{\citenamefont {Engblom}\ \emph {et~al.}(2009)\citenamefont
  {Engblom}, \citenamefont {Ferm}, \citenamefont {Hellander},\ and\
  \citenamefont {L\"otstedt}}]{Engblom:2009:SSR}%
  \BibitemOpen
  \bibfield  {author} {\bibinfo {author} {\bibfnamefont {S.}~\bibnamefont
  {Engblom}}, \bibinfo {author} {\bibfnamefont {L.}~\bibnamefont {Ferm}},
  \bibinfo {author} {\bibfnamefont {A.}~\bibnamefont {Hellander}}, \ and\
  \bibinfo {author} {\bibfnamefont {P.}~\bibnamefont {L\"otstedt}},\
  }\href@noop {} {\bibfield  {journal} {\bibinfo  {journal} {SIAM Journal on
  Scientific Computing}\ }\textbf {\bibinfo {volume} {31}},\ \bibinfo {pages}
  {1774} (\bibinfo {year} {2009})}%\BibitemShut {NoStop}%
\bibitem [{\citenamefont {Erban}\ and\ \citenamefont
  {Chapman}(2009)}]{Erban:2009:SMR}%
  \BibitemOpen
  \bibfield  {author} {\bibinfo {author} {\bibfnamefont {R.}~\bibnamefont
  {Erban}}\ and\ \bibinfo {author} {\bibfnamefont {S.~J.}\ \bibnamefont
  {Chapman}},\ }\href@noop {} {\bibfield  {journal} {\bibinfo  {journal}
  {Physical Biology}\ }\textbf {\bibinfo {volume} {6}},\ \bibinfo {pages}
  {046001} (\bibinfo {year} {2009})}%\BibitemShut {NoStop}%
\bibitem [{\citenamefont {van Gunsteren}\ and\ \citenamefont
  {Berendsen}(1982)}]{vanGunsteren:1982:ABD}%
  \BibitemOpen
  \bibfield  {author} {\bibinfo {author} {\bibfnamefont {W.}~\bibnamefont {van
  Gunsteren}}\ and\ \bibinfo {author} {\bibfnamefont {H.}~\bibnamefont
  {Berendsen}},\ }\href@noop {} {\bibfield  {journal} {\bibinfo  {journal}
  {Molecular Physics}\ }\textbf {\bibinfo {volume} {45}},\ \bibinfo {pages}
  {637} (\bibinfo {year} {1982})}%\BibitemShut {NoStop}%
\bibitem [{\citenamefont {Smoluchowski}(1917)}]{Smoluchowski:1917:VMT}%
  \BibitemOpen
  \bibfield  {author} {\bibinfo {author} {\bibfnamefont {M.}~\bibnamefont
  {Smoluchowski}},\ }\href@noop {} {\bibfield  {journal} {\bibinfo  {journal}
  {Zeitschrift f\"ur physikalische Chemie}\ }\textbf {\bibinfo {volume} {92}},\
  \bibinfo {pages} {129} (\bibinfo {year} {1917})}%\BibitemShut {NoStop}%
\bibitem [{\citenamefont {Erban}, \citenamefont {Chapman},\ and\ \citenamefont
  {Maini}(2007)}]{Erban:2007:PGS}%
  \BibitemOpen
  \bibfield  {author} {\bibinfo {author} {\bibfnamefont {R.}~\bibnamefont
  {Erban}}, \bibinfo {author} {\bibfnamefont {S.~J.}\ \bibnamefont {Chapman}},
  \ and\ \bibinfo {author} {\bibfnamefont {P.}~\bibnamefont {Maini}},\
  }\href@noop {} {\enquote {\bibinfo {title} {A practical guide to stochastic
  simulations of reaction-diffusion processes},}\ } (\bibinfo {year} {2007}),\
  \bibinfo {note} {35 pages, available as
  http://arxiv.org/abs/0704.1908}%\BibitemShut {NoStop}%
\bibitem [{\citenamefont {Fisher}(1937)}]{Fisher:1937:WAA}%
  \BibitemOpen
  \bibfield  {author} {\bibinfo {author} {\bibfnamefont {R.}~\bibnamefont
  {Fisher}},\ }\href@noop {} {\bibfield  {journal} {\bibinfo  {journal} {Annals
  of Eugenics}\ }\textbf {\bibinfo {volume} {7}},\ \bibinfo {pages} {355}
  (\bibinfo {year} {1937})}%\BibitemShut {NoStop}%
\bibitem [{\citenamefont {Panja}(2004)}]{Panja:2004:EFP}%
  \BibitemOpen
  \bibfield  {author} {\bibinfo {author} {\bibfnamefont {D.}~\bibnamefont
  {Panja}},\ }\href@noop {} {\bibfield  {journal} {\bibinfo  {journal} {Physics
  Reports}\ }\textbf {\bibinfo {volume} {393}},\ \bibinfo {pages} {87}
  (\bibinfo {year} {2004})}%\BibitemShut {NoStop}%
\bibitem [{\citenamefont {van Saarloos}(2003)}]{vanSaarloos:2003:FPU}%
  \BibitemOpen
  \bibfield  {author} {\bibinfo {author} {\bibfnamefont {W.}~\bibnamefont {van
  Saarloos}},\ }\href@noop {} {\bibfield  {journal} {\bibinfo  {journal}
  {Physics Reports}\ }\textbf {\bibinfo {volume} {386}},\ \bibinfo {pages} {29}
  (\bibinfo {year} {2003})}%\BibitemShut {NoStop}%
\bibitem [{\citenamefont {Brunet}\ and\ \citenamefont
  {Derrida}(2001)}]{Brunet:2001:EMN}%
  \BibitemOpen
  \bibfield  {author} {\bibinfo {author} {\bibfnamefont {{\'E}.}~\bibnamefont
  {Brunet}}\ and\ \bibinfo {author} {\bibfnamefont {B.}~\bibnamefont
  {Derrida}},\ }\href@noop {} {\bibfield  {journal} {\bibinfo  {journal}
  {Journal of Statistical Physics}\ }\textbf {\bibinfo {volume} {103}},\
  \bibinfo {pages} {269} (\bibinfo {year} {2001})}%\BibitemShut {NoStop}%
\bibitem [{\citenamefont {Breuer}, \citenamefont {Huber},\ and\ \citenamefont
  {Petruccione}(1994)}]{Breuer:1994:FEW}%
  \BibitemOpen
  \bibfield  {author} {\bibinfo {author} {\bibfnamefont {H.}~\bibnamefont
  {Breuer}}, \bibinfo {author} {\bibfnamefont {W.}~\bibnamefont {Huber}}, \
  and\ \bibinfo {author} {\bibfnamefont {F.}~\bibnamefont {Petruccione}},\
  }\href@noop {} {\bibfield  {journal} {\bibinfo  {journal} {Physica D:
  Nonlinear Phenomena}\ }\textbf {\bibinfo {volume} {73}},\ \bibinfo {pages}
  {259} (\bibinfo {year} {1994})}%\BibitemShut {NoStop}%
\bibitem [{\citenamefont {Moro}(2004)}]{Moro:2004:HMS}%
  \BibitemOpen
  \bibfield  {author} {\bibinfo {author} {\bibfnamefont {E.}~\bibnamefont
  {Moro}},\ }\href@noop {} {\bibfield  {journal} {\bibinfo  {journal} {Physical
  Review E}\ }\textbf {\bibinfo {volume} {69}},\ \bibinfo {pages} {060101}
  (\bibinfo {year} {2004})}%\BibitemShut {NoStop}%
\bibitem [{\citenamefont {Flegg}, \citenamefont {Chapman},\ and\ \citenamefont
  {Erban}(2012)}]{Flegg:2012:TRM}%
  \BibitemOpen
  \bibfield  {author} {\bibinfo {author} {\bibfnamefont {M.}~\bibnamefont
  {Flegg}}, \bibinfo {author} {\bibfnamefont {J.}~\bibnamefont {Chapman}}, \
  and\ \bibinfo {author} {\bibfnamefont {R.}~\bibnamefont {Erban}},\
  }\href@noop {} {\bibfield  {journal} {\bibinfo  {journal} {Journal of the
  Royal Society Interface}\ }\textbf {\bibinfo {volume} {9}},\ \bibinfo {pages}
  {859} (\bibinfo {year} {2012})}%\BibitemShut {NoStop}%
\bibitem [{\citenamefont {Flegg}\ \emph {et~al.}(2013)\citenamefont {Flegg},
  \citenamefont {Chapman}, \citenamefont {Zheng},\ and\ \citenamefont
  {Erban}}]{Flegg:2013:ATM}%
  \BibitemOpen
  \bibfield  {author} {\bibinfo {author} {\bibfnamefont {M.}~\bibnamefont
  {Flegg}}, \bibinfo {author} {\bibfnamefont {J.}~\bibnamefont {Chapman}},
  \bibinfo {author} {\bibfnamefont {L.}~\bibnamefont {Zheng}}, \ and\ \bibinfo
  {author} {\bibfnamefont {R.}~\bibnamefont {Erban}},\ }\href@noop {} {\enquote
  {\bibinfo {title} {Analysis of the two-regime method on square meshes},}\ }
  (\bibinfo {year} {2013}),\ \bibinfo {note} {submitted to {\it SIAM Journal on
  Scientific Computing}}%\BibitemShut {NoStop}%
\bibitem [{\citenamefont {Erban}, \citenamefont {Flegg},\ and\ \citenamefont
  {Papoian}(2013)}]{Erban:2013:MSR}%
  \BibitemOpen
  \bibfield  {author} {\bibinfo {author} {\bibfnamefont {R.}~\bibnamefont
  {Erban}}, \bibinfo {author} {\bibfnamefont {M.}~\bibnamefont {Flegg}}, \ and\
  \bibinfo {author} {\bibfnamefont {G.}~\bibnamefont {Papoian}},\ }\href@noop
  {} {\bibfield  {journal} {\bibinfo  {journal} {Bulletin of Mathematical
  Biology}\ }\textbf {\bibinfo {volume} {to appear}},\ \bibinfo {pages} {DOI:
  10.1007/s11538} (\bibinfo {year} {2013})}%\BibitemShut {NoStop}%
\bibitem [{\citenamefont {Andrews}(2012)}]{Andrews:2012:SSC}%
  \BibitemOpen
  \bibfield  {author} {\bibinfo {author} {\bibfnamefont {S.}~\bibnamefont
  {Andrews}},\ }in\ \href@noop {} {\emph {\bibinfo {booktitle} {Bacterial
  Molecular Networks}}}\ (\bibinfo  {publisher} {Springer},\ \bibinfo {year}
  {2012})\ pp.\ \bibinfo {pages} {519--542}%\BibitemShut {NoStop}%
\bibitem [{\citenamefont {Stiles}\ and\ \citenamefont
  {Bartol}(2001)}]{Stiles:2001:MCM}%
  \BibitemOpen
  \bibfield  {author} {\bibinfo {author} {\bibfnamefont {J.}~\bibnamefont
  {Stiles}}\ and\ \bibinfo {author} {\bibfnamefont {T.}~\bibnamefont
  {Bartol}},\ }in\ \href@noop {} {\emph {\bibinfo {booktitle} {Computational
  Neuroscience: Realistic Modeling for Experimentalists}}},\ \bibinfo {editor}
  {edited by\ \bibinfo {editor} {\bibfnamefont {E.}~\bibnamefont {Schutter}}}\
  (\bibinfo  {publisher} {CRC Press},\ \bibinfo {year} {2001})\ pp.\ \bibinfo
  {pages} {87--127}%\BibitemShut {NoStop}%
\bibitem [{\citenamefont {Kerr}\ \emph {et~al.}(2008)\citenamefont {Kerr},
  \citenamefont {Bartol}, \citenamefont {Kaminsky}, \citenamefont {Dittrich},
  \citenamefont {Chang}, \citenamefont {Baden}, \citenamefont {Sejnowski},\
  and\ \citenamefont {Stiles}}]{Kerr:2008:FMC}%
  \BibitemOpen
  \bibfield  {author} {\bibinfo {author} {\bibfnamefont {R.}~\bibnamefont
  {Kerr}}, \bibinfo {author} {\bibfnamefont {T.}~\bibnamefont {Bartol}},
  \bibinfo {author} {\bibfnamefont {B.}~\bibnamefont {Kaminsky}}, \bibinfo
  {author} {\bibfnamefont {M.}~\bibnamefont {Dittrich}}, \bibinfo {author}
  {\bibfnamefont {J.}~\bibnamefont {Chang}}, \bibinfo {author} {\bibfnamefont
  {S.}~\bibnamefont {Baden}}, \bibinfo {author} {\bibfnamefont
  {T.}~\bibnamefont {Sejnowski}}, \ and\ \bibinfo {author} {\bibfnamefont
  {J.}~\bibnamefont {Stiles}},\ }\href@noop {} {\bibfield  {journal} {\bibinfo
  {journal} {SIAM Journal on Scientific Computing}\ }\textbf {\bibinfo {volume}
  {30}},\ \bibinfo {pages} {3126} (\bibinfo {year} {2008})}%\BibitemShut
  {NoStop}%
\bibitem [{\citenamefont {Drawert}, \citenamefont {Engblom},\ and\
  \citenamefont {Hellander}(2012)}]{Drawert:2012:UMF}%
  \BibitemOpen
  \bibfield  {author} {\bibinfo {author} {\bibfnamefont {B.}~\bibnamefont
  {Drawert}}, \bibinfo {author} {\bibfnamefont {S.}~\bibnamefont {Engblom}}, \
  and\ \bibinfo {author} {\bibfnamefont {A.}~\bibnamefont {Hellander}},\
  }\href@noop {} {\bibfield  {journal} {\bibinfo  {journal} {BMC Systems
  Biology}\ }\textbf {\bibinfo {volume} {6}},\ \bibinfo {pages} {76} (\bibinfo
  {year} {2012})}%\BibitemShut {NoStop}%
\bibitem [{\citenamefont {Elf}\ and\ \citenamefont
  {Ehrenberg}(2004)}]{Elf:2004:SSB}%
  \BibitemOpen
  \bibfield  {author} {\bibinfo {author} {\bibfnamefont {J.}~\bibnamefont
  {Elf}}\ and\ \bibinfo {author} {\bibfnamefont {M.}~\bibnamefont
  {Ehrenberg}},\ }\href@noop {} {\bibfield  {journal} {\bibinfo  {journal}
  {Systems biology}\ }\textbf {\bibinfo {volume} {1}},\ \bibinfo {pages} {230}
  (\bibinfo {year} {2004})}%\BibitemShut {NoStop}%
\bibitem [{\citenamefont {Gillespie}(1977)}]{Gillespie:1977:ESS}%
  \BibitemOpen
  \bibfield  {author} {\bibinfo {author} {\bibfnamefont {D.}~\bibnamefont
  {Gillespie}},\ }\href@noop {} {\bibfield  {journal} {\bibinfo  {journal}
  {Journal of Physical Chemistry}\ }\textbf {\bibinfo {volume} {81}},\ \bibinfo
  {pages} {2340} (\bibinfo {year} {1977})}%\BibitemShut {NoStop}%
\bibitem [{\citenamefont {Gibson}\ and\ \citenamefont
  {Bruck}(2000)}]{Gibson:2000:EES}%
  \BibitemOpen
  \bibfield  {author} {\bibinfo {author} {\bibfnamefont {M.}~\bibnamefont
  {Gibson}}\ and\ \bibinfo {author} {\bibfnamefont {J.}~\bibnamefont {Bruck}},\
  }\href@noop {} {\bibfield  {journal} {\bibinfo  {journal} {Journal of
  Physical Chemistry A}\ }\textbf {\bibinfo {volume} {104}},\ \bibinfo {pages}
  {1876} (\bibinfo {year} {2000})}%\BibitemShut {NoStop}%
\bibitem [{\citenamefont {Erban}\ and\ \citenamefont
  {Chapman}(2007)}]{Erban:2007:RBC}%
  \BibitemOpen
  \bibfield  {author} {\bibinfo {author} {\bibfnamefont {R.}~\bibnamefont
  {Erban}}\ and\ \bibinfo {author} {\bibfnamefont {S.~J.}\ \bibnamefont
  {Chapman}},\ }\href@noop {} {\bibfield  {journal} {\bibinfo  {journal}
  {Physical Biology}\ }\textbf {\bibinfo {volume} {4}},\ \bibinfo {pages} {16}
  (\bibinfo {year} {2007})}%\BibitemShut {NoStop}%
\bibitem [{\citenamefont {Hattne}, \citenamefont {Fange},\ and\ \citenamefont
  {Elf}(2005)}]{Hattne:2005:SRD}%
  \BibitemOpen
  \bibfield  {author} {\bibinfo {author} {\bibfnamefont {J.}~\bibnamefont
  {Hattne}}, \bibinfo {author} {\bibfnamefont {D.}~\bibnamefont {Fange}}, \
  and\ \bibinfo {author} {\bibfnamefont {J.}~\bibnamefont {Elf}},\ }\href@noop
  {} {\bibfield  {journal} {\bibinfo  {journal} {Bioinfor\-matics}\ }\textbf
  {\bibinfo {volume} {21}},\ \bibinfo {pages} {2923} (\bibinfo {year}
  {2005})}%\BibitemShut {NoStop}%
\bibitem [{\citenamefont {Flegg}, \citenamefont {R\"udiger},\ and\
  \citenamefont {Erban}(2013)}]{Flegg:2013:DSN}%
  \BibitemOpen
  \bibfield  {author} {\bibinfo {author} {\bibfnamefont {M.}~\bibnamefont
  {Flegg}}, \bibinfo {author} {\bibfnamefont {S.}~\bibnamefont {R\"udiger}}, \
  and\ \bibinfo {author} {\bibfnamefont {R.}~\bibnamefont {Erban}},\
  }\href@noop {} {\bibfield  {journal} {\bibinfo  {journal} {Journal of
  Chemical Physics}\ }\textbf {\bibinfo {volume} {138}},\ \bibinfo {pages}
  {154103} (\bibinfo {year} {2013})}%\BibitemShut {NoStop}%
\bibitem [{\citenamefont {Ho}(2012)}]{Ho:2012:MSR}%
  \BibitemOpen
  \bibfield  {author} {\bibinfo {author} {\bibfnamefont {C.-P.}\ \bibnamefont
  {Ho}},\ }\emph {\bibinfo {title} {Multi-scale reaction diffusion simulations
  in biology}},\ \href@noop {} {\bibinfo {type} {{M.S}c. {T}hesis}},\ \bibinfo
  {school} {University of Oxford} (\bibinfo {year} {2012})%\BibitemShut{NoStop}%
\bibitem [{\citenamefont {Tostevin}, \citenamefont {ten Wolde},\ and\
  \citenamefont {Howard}(2007)}]{Tostevin:2007:FLP}%
  \BibitemOpen
  \bibfield  {author} {\bibinfo {author} {\bibfnamefont {F.}~\bibnamefont
  {Tostevin}}, \bibinfo {author} {\bibfnamefont {P.}~\bibnamefont {ten Wolde}},
  \ and\ \bibinfo {author} {\bibfnamefont {M.}~\bibnamefont {Howard}},\
  }\href@noop {} {\bibfield  {journal} {\bibinfo  {journal} {PLOS Computational
  Biology}\ }\textbf {\bibinfo {volume} {3}},\ \bibinfo {pages} {763} (\bibinfo
  {year} {2007})}%\BibitemShut {NoStop}%
\bibitem [{\citenamefont {Howard}(2012)}]{Howard:2012:HBR}%
  \BibitemOpen
  \bibfield  {author} {\bibinfo {author} {\bibfnamefont {M.}~\bibnamefont
  {Howard}},\ }\href@noop {} {\bibfield  {journal} {\bibinfo  {journal} {Trends
  in Cell Biology}\ }\textbf {\bibinfo {volume} {22}},\ \bibinfo {pages} {311}
  (\bibinfo {year} {2012})}%\BibitemShut {NoStop}%
\bibitem [{\citenamefont {Bergmann}\ \emph {et~al.}(2007)\citenamefont
  {Bergmann}, \citenamefont {Sandler}, \citenamefont {Sberro}, \citenamefont
  {Shnider}, \citenamefont {Schejter}, \citenamefont {Shilo},\ and\
  \citenamefont {Barkai}}]{Bergmann:2007:PDB}%
  \BibitemOpen
  \bibfield  {author} {\bibinfo {author} {\bibfnamefont {S.}~\bibnamefont
  {Bergmann}}, \bibinfo {author} {\bibfnamefont {O.}~\bibnamefont {Sandler}},
  \bibinfo {author} {\bibfnamefont {H.}~\bibnamefont {Sberro}}, \bibinfo
  {author} {\bibfnamefont {S.}~\bibnamefont {Shnider}}, \bibinfo {author}
  {\bibfnamefont {E.}~\bibnamefont {Schejter}}, \bibinfo {author}
  {\bibfnamefont {B.}~\bibnamefont {Shilo}}, \ and\ \bibinfo {author}
  {\bibfnamefont {N.}~\bibnamefont {Barkai}},\ }\href@noop {} {\bibfield
  {journal} {\bibinfo  {journal} {PLoS Biology}\ }\textbf {\bibinfo {volume}
  {5}},\ \bibinfo {pages} {e46} (\bibinfo {year} {2007})}%\BibitemShut {NoStop}%
\bibitem [{\citenamefont {Flegg}, \citenamefont {Hellander},\ and\
  \citenamefont {Erban}(2013)}]{Flegg:2013:CMC}%
  \BibitemOpen
  \bibfield  {author} {\bibinfo {author} {\bibfnamefont {M.}~\bibnamefont
  {Flegg}}, \bibinfo {author} {\bibfnamefont {S.}~\bibnamefont {Hellander}}, \
  and\ \bibinfo {author} {\bibfnamefont {R.}~\bibnamefont {Erban}},\
  }\href@noop {} {\enquote {\bibinfo {title} {Convergence of methods for
  coupling of microscopic and mesoscopic reaction-diffusion simulations},}\ }
  (\bibinfo {year} {2013}),\ \bibinfo {note} {submitted to {\it Journal of
  Computational Physics}}%\BibitemShut {NoStop}%
\bibitem [{\citenamefont {Alexander}, \citenamefont {Garcia},\ and\
  \citenamefont {Tartakovsky}(2002)}]{Alexander:2002:ARS}%
  \BibitemOpen
  \bibfield  {author} {\bibinfo {author} {\bibfnamefont {F.}~\bibnamefont
  {Alexander}}, \bibinfo {author} {\bibfnamefont {A.}~\bibnamefont {Garcia}}, \
  and\ \bibinfo {author} {\bibfnamefont {D.}~\bibnamefont {Tartakovsky}},\
  }\href@noop {} {\bibfield  {journal} {\bibinfo  {journal} {Journal of
  Computational Physics}\ }\textbf {\bibinfo {volume} {182}},\ \bibinfo {pages}
  {47} (\bibinfo {year} {2002})}%\BibitemShut {NoStop}%
\bibitem [{\citenamefont {Geyer}, \citenamefont {Gorba},\ and\ \citenamefont
  {Helms}(2004)}]{Geyer:2004:IBD}%
  \BibitemOpen
  \bibfield  {author} {\bibinfo {author} {\bibfnamefont {T.}~\bibnamefont
  {Geyer}}, \bibinfo {author} {\bibfnamefont {C.}~\bibnamefont {Gorba}}, \ and\
  \bibinfo {author} {\bibfnamefont {V.}~\bibnamefont {Helms}},\ }\href@noop {}
  {\bibfield  {journal} {\bibinfo  {journal} {Journal of Chemical Physics}\
  }\textbf {\bibinfo {volume} {120}},\ \bibinfo {pages} {4573} (\bibinfo {year}
  {2004})}%\BibitemShut {NoStop}%
\bibitem [{\citenamefont {Wagner}\ and\ \citenamefont
  {Flekk{\o}y}(2004)}]{Wagner:2004:HCF}%
  \BibitemOpen
  \bibfield  {author} {\bibinfo {author} {\bibfnamefont {G.}~\bibnamefont
  {Wagner}}\ and\ \bibinfo {author} {\bibfnamefont {E.}~\bibnamefont
  {Flekk{\o}y}},\ }\href@noop {} {\bibfield  {journal} {\bibinfo  {journal}
  {Philosophical Transactions of the Royal Society A: Mathematical, Physical \&
  Engineering Sciences}\ }\textbf {\bibinfo {volume} {362}},\ \bibinfo {pages}
  {1655} (\bibinfo {year} {2004})}%\BibitemShut {NoStop}%
\bibitem [{\citenamefont {Franz}\ \emph {et~al.}(2013)\citenamefont {Franz},
  \citenamefont {Flegg}, \citenamefont {Chapman},\ and\ \citenamefont
  {Erban}}]{Franz:2013:MRA}%
  \BibitemOpen
  \bibfield  {author} {\bibinfo {author} {\bibfnamefont {B.}~\bibnamefont
  {Franz}}, \bibinfo {author} {\bibfnamefont {M.}~\bibnamefont {Flegg}},
  \bibinfo {author} {\bibfnamefont {J.}~\bibnamefont {Chapman}}, \ and\
  \bibinfo {author} {\bibfnamefont {R.}~\bibnamefont {Erban}},\ }\href@noop {}
  {\bibfield  {journal} {\bibinfo  {journal} {SIAM Journal on Applied
  Mathematics}\ }\textbf {\bibinfo {volume} {73}},\ \bibinfo {pages} {1224}
  (\bibinfo {year} {2013})}%\BibitemShut {NoStop}%
\end{thebibliography}
\end{document}